\documentclass[trackchanges]{aastex7}
\graphicspath{{./}{figures/}}
\usepackage[ruled,linesnumbered]{algorithm2e}
\usepackage{algpseudocode}
\usepackage{amsmath}
\usepackage{xcolor}

\begin{document}

\title{Study of the IC 443 region with the HAWC observatory}

\author[0000-0001-8749-1647]{R.~Alfaro}
\affiliation{Instituto de F\'{i}sica, Universidad Nacional Aut\'{o}noma de M\'{e}xico, Ciudad de M\'{e}xico, M\'{e}xico }
\email{ruben@fisica.unam.mx}

\author{C.~Alvarez}
\affiliation{Universidad Aut\'{o}noma de Chiapas, Tuxtla Guti\'{e}rrez, Chiapas, M\'{e}xico}
\email{crabpulsar@hotmail.com}

\author[0000-0002-0595-9267]{M.~Araya}
\affiliation{Escuela de F\'{i}sica, Universidad de Costa Rica, San Jos\'{e}, Costa Rica}
\email{miguel.araya@ucr.ac.cr}

\author{J.C.~Arteaga-Vel\'{a}zquez}
\affiliation{Universidad Michoacana de San Nicol\'{a}s de Hidalgo, Morelia, M\'{e}xico }
\email{juan.arteaga@umich.mx}

\author[0000-0002-4020-4142]{D.~Avila Rojas}
\affiliation{Instituto de F\'{i}sica, Universidad Nacional Aut\'{o}noma de M\'{e}xico, Ciudad de M\'{e}xico, M\'{e}xico }
\email{doavila@astro.unam.mx}

\author[0000-0002-2084-5049]{H.A.~Ayala Solares}
\affiliation{Department of Physics, Pennsylvania State University, University Park, PA, USA }
\email[show]{hgayala@psu.edu}

\author[0000-0002-5529-6780]{R.~Babu}
\affiliation{Department of Physics and Astronomy, Michigan State University, East Lansing, MI, US}
\email{baburish@msu.edu}

\author[0000-0002-3886-3739]{P.~Bangale}
\affiliation{College of Science and Technology, Temple University, Philadelphia, PA, US}
\email{priyadarshini.bangale@temple.edu}

\author{A.~Bernal}
\affiliation{Instituto de Astronom\'{i}a, Universidad Nacional Autónoma de México, Ciudad de Mexico, Mexico} 
\email{abel@astro.unam.mx}

\author[0000-0002-4042-3855]{K.S.~Caballero-Mora}
\affiliation{Universidad Autónoma de Chiapas, Tuxtla Gutiérrez, Chiapas, México}
\email{karen.scm@gmail.com}

\author[0000-0003-2158-2292]{T.~Capistran} 
\affiliation{Università degli Studi di Torino, I-10125 Torino, Italy} 
\email{tcapistranc@gmail.com}

\author[0000-0002-8553-3302]{A.~Carrami\~{n}ana}
\affiliation{Instituto Nacional de Astrof\'{i}sica, \'{O}ptica y Electr\'{o}nica, Puebla, M\'{e}xico }
\email{alberto@inaoep.mx}

\author[0000-0002-6144-9122]{S.~Casanova}
\affiliation{Institute of Nuclear Physics Polish Academy of Sciences, PL-31342 IFJ-PAN, Krakow, Poland }
\email{sabrinacasanova@gmail.com}

\author[0000-0002-7607-9582]{U.~Cotti}
\affiliation{Universidad Michoacana de San Nicol\'{a}s de Hidalgo, Morelia, M\'{e}xico }
\email{umberto.cotti@umich.mx}

\author[0000-0002-1132-871X]{J.~Cotzomi}
\affiliation{Facultad de Ciencias F\'{i}sico Matemáticas, Benem\'{e}rita Universidad Aut\'{o}noma de Puebla, Puebla, M\'{e}xico }
\email{jcotzomi@yahoo.com.mx}

\author[0000-0002-7747-754X]{S.~Couti\~no de Le\'{o}n}
\affiliation{Department of Physics, University of Wisconsin-Madison, Madison, WI 53706, USA }
\email{sara.cdl989@gmail.com}

\author[0000-0001-9643-4134]{E.~De la Fuente}
\affiliation{Departamento de F\'{i}sica, Centro Universitario de Ciencias Exactas e Ingenierias, Universidad de Guadalajara, Guadalajara, M\'{e}xico }
\email{edfuente@gmail.com}

\author[0000-0002-2672-4141]{D.~Depaoli}
\affiliation{Max-Planck Institute for Nuclear Physics, 69117 Heidelberg, Germany} 
\email{davide.depaoli@mpi-hd.mpg.de}

\author{P.~Desiati} 
\affiliation{Department of Physics, University of Wisconsin-Madison, Madison, WI 53706, USA }
\email{paolo.desiati@icecube.wisc.edu}

\author[0000-0002-7574-1298]{N.~Di Lalla}
\affiliation{Department of Physics, Stanford University: Stanford, CA 94305–4060, USA}
\email{niccolo.dilalla@stanford.edu}

\author{R.~Diaz Hernandez}
\affiliation{Instituto Nacional de Astrof\'{i}sica, \'{O}ptica y Electr\'{o}nica, Puebla, M\'{e}xico }
\email{dihera77@gmail.com}

\author[0000-0001-8451-7450]{B.L.~Dingus}
\affiliation{Physics Division, Los Alamos National Laboratory, Los Alamos, NM, USA }
\email{dingus@lanl.gov}

\author[0000-0002-2987-9691]{M.A.~DuVernois}
\affiliation{Department of Physics, University of Wisconsin-Madison, Madison, WI 53706, USA }
\email{duvernois@icecube.wisc.edu}

\author[0000-0002-0087-0693]{J.C.~Díaz-Vélez}
\affiliation{Department of Physics, University of Wisconsin-Madison, Madison, WI 53706, USA }
\email{juancarlos@icecube.wisc.edu}

\author[0000-0003-2423-4656]{T.~Ergin}
\affiliation{Department of Physics and Astronomy, Michigan State University, East Lansing, MI, USA}
\email{ergintul@msu.edu}

\author[0000-0001-7074-1726]{C.~Espinoza}
\affiliation{Instituto de F\'{i}sica, Universidad Nacional Aut\'{o}noma de M\'{e}xico, Ciudad de M\'{e}xico, M\'{e}xico }
\email{m.catalina@fisica.unam.mx}

\author[0000-0002-5387-8138]{K.~Fang}
\affiliation{Department of Physics, University of Wisconsin-Madison, Madison, WI 53706, USA }
\email{kefang@physics.wisc.edu}

\author[0000-0002-0173-6453]{N.~Fraija}
\affiliation{Instituto de Astronom\'{i}a, Universidad Nacional Aut\'{o}noma de M\'{e}xico, Ciudad de M\'{e}xico, M\'{e}xico }
\email{nifraija@astro.unam.mx}

\author{S.~Fraija}
\affiliation{Instituto de Astronom\'{i}a, Universidad Nacional Autónoma de México, Ciudad de Mexico, Mexico}
\email{sarafraija@hotmail.com}

\author[0000-0002-4188-5584]{J.A.~García-González}
\affiliation{Tecnologico de Monterrey, Escuela de Ingenier\'{i}a y Ciencias, Ave. Eugenio Garza Sada 2501, Monterrey, N.L., Mexico, 64849}
\email{anteus79@tec.mx}

\author{H.~Goksu}
\affiliation{Max-Planck Institute for Nuclear Physics, 69117 Heidelberg, Germany}
\email{hazal.goksu@mpi-hd.mpg.de}

\author{J.A.~González-Cervera}
\affiliation{Universidad Michoacana de San Nicolás de Hidalgo, Morelia, Mexico}
\email{jose.gonzalez.c@umich.mx}

\author[0000-0002-5209-5641]{M.M.~Gonz\'{a}lez}
\affiliation{Instituto de Astronom\'{i}a, Universidad Nacional Aut\'{o}noma de M\'{e}xico, Ciudad de M\'{e}xico, M\'{e}xico }
\email{magda@astro.unam.mx}

\author[0000-0002-9790-1299]{J.A.~Goodman}
\affiliation{Dept. of Physics, University of Maryland, College Park, MD 20742, USA}
\email{goodman@umd.edu}

\author{S.~Groetsch} 
\affiliation{Department of Physics, Michigan Technological University, Houghton, MI, USA}
\email{sjgroets@mtu.edu}

\author[0000-0001-9844-2648]{J.P.~Harding}
\affiliation{Physics Division, Los Alamos National Laboratory, Los Alamos, NM, USA }
\email{jpharding@lanl.gov}

\author{S.~Hernández-Cadena}
\affiliation{Tsung-Dao Lee Institute \& School of Physics and Astronomy, Shanghai Jiao Tong University, Shanghai, China}
\email{shkdna@sjtu.edu.cn}

\author[0000-0001-5169-723X]{I.~Herzog}
\affiliation{Department of Physics and Astronomy, Michigan State University, East Lansing, MI, USA}
\email{herzogia@msu.edu}

\author[0000-0002-1031-7760]{J.~Hinton} 
\affiliation{Max-Planck Institute for Nuclear Physics, 69117 Heidelberg, Germany}
\email{jim.hinton@mpi-hd.mpg.de}

\author[0000-0002-3808-4639]{D.~Huang} 
\affiliation{Department of Physics, University of Maryland, College Park, MD, USA}
\email{dezhih@mtu.edu}

\author[0000-0002-5527-7141]{F.~Hueyotl-Zahuantitla}
\affiliation{Universidad Aut\'{o}noma de Chiapas, Tuxtla Guti\'{e}rrez, Chiapas, M\'{e}xico}
\email{filihz@gmail.com}

\author[0000-0002-3302-7897]{P.~H{\"u}ntemeyer}
\affiliation{Department of Physics, Michigan Technological University, Houghton, MI, USA }
\email{petra@mtu.edu}

\author{S.~Kaufmann} 
\affiliation{Universidad Politecnica de Pachuca, Pachuca, Hgo, M\'{e}xico }
\email{skaufmann13@googlemail.com}

\author[0000-0001-6336-5291]	{A.~Lara}
\affiliation{Instituto de Geof\'{i}sica, Universidad Nacional Aut\'{o}noma de M\'{e}xico, Ciudad de M\'{e}xico, M\'{e}xico }
\email{alara@igeofisica.unam.mx}

\author[0000-0002-2153-1519]{J.~Lee}
\affiliation{University of Seoul, Seoul, Rep. of Korea}
\email{jason.lee@uos.ac.kr}

\author[0000-0001-5516-4975]{H.~Le\'{o}n Vargas}
\affiliation{Instituto de F\'{i}sica, Universidad Nacional Aut\'{o}noma de M\'{e}xico, Ciudad de M\'{e}xico, M\'{e}xico }
\email{hleonvar@fisica.unam.mx}

\author[0000-0003-2696-947X]{J.T.~Linnemann}
\affiliation{Dept. of Physics and Astronomy, Michigan State University, East Lansing, MI 48824, USA}
\email{linneman@msu.edu}

\author[0000-0001-8825-3624]{A.L.~Longinotti}
\affiliation{Instituto Nacional de Astrof\'{i}sica, \'{O}ptica y Electr\'{o}nica, Puebla, M\'{e}xico }
\email{alonginotti@astro.unam.mx}

\author[0000-0003-2810-4867]{G.~Luis-Raya}
\affiliation{Universidad Politecnica de Pachuca, Pachuca, Hgo, M\'{e}xico }
\email{gilura6969@hotmail.com}

\author[0000-0001-8088-400X]{K.~Malone}
\affiliation{Space Science and Applications Group, Los Alamos National Laboratory, Los Alamos, NM, USA }
\email{kmalone@lanl.gov}

\author[0000-0001-9052-856X]{O.~Martinez}
\affiliation{Facultad de Ciencias F\'{i}sico Matemáticas, Benem\'{e}rita Universidad Aut\'{o}noma de Puebla, Puebla, M\'{e}xico }
\email{omartin@fcfm.buap.mx}

\author[0000-0002-2824-3544]{J.~Mart\'{i}nez-Castro}
\affiliation{Centro de Investigaci\'on en Computaci\'on, Instituto Polit\'ecnico Nacional, Mexico City, Mexico.}
\email{macj@cic.ipn.mx}

\author[0000-0002-2610-863X]{J.A.~Matthews}
\affiliation{Dept of Physics and Astronomy, University of New Mexico, Albuquerque, NM, USA }
\email{johnm@unm.edu}

\author[0000-0002-8390-9011]{P.~Miranda-Romagnoli}
\affiliation{Universidad Aut\'{o}noma del Estado de Hidalgo, Pachuca, M\'{e}xico }
\email{pa.miranda.r@gmail.com}

\author{J.A.~Montes}
\affiliation{Instituto de Astronom\'{i}a, Universidad Nacional Autónoma de México, Ciudad de Mexico, Mexico}
\email{jamontes@astro.unam.mx}

\author[0000-0001-9361-0147]{J.A.~Morales-Soto}
\affiliation{Universidad Michoacana de San Nicol\'{a}s de Hidalgo, Morelia, M\'{e}xico }
\email{jmoralessg@gmail.com}

\author[0000-0002-1114-2640]{E.~Moreno}
\affiliation{Facultad de Ciencias F\'{i}sico Matemáticas, Benem\'{e}rita Universidad Aut\'{o}noma de Puebla, Puebla, M\'{e}xico }
\email{emoreno@fcfm.buap.mx}

\author[0000-0002-7675-4656]{M.~Mostaf\'{a}}
\affiliation{College of Science and Technology, Temple University, Philadelphia, PA, US}
\email{miguel.mostafa@temple.edu}

\author{M.~Najafi}
\affiliation{Department of Physics, Michigan Technological University, Houghton, MI, USA}
\email{mnajafi@mtu.edu}

\author[0000-0003-1059-8731]{L.~Nellen}
\affiliation{Instituto de Ciencias Nucleares, Universidad Nacional Aut\'{o}noma de M\'{e}xico, Ciudad de M\'{e}xico, M\'{e}xico }
\email{lukas@nucleares.unam.mx}

\author[0000-0002-6859-3944]{M.U.~Nisa}
\affiliation{Dept. of Physics and Astronomy, Michigan State University, East Lansing, MI 48824, USA}
\email{nisamehr@msu.edu}

\author[0000-0001-7099-108X]{R.~Noriega-Papaqui}
\affiliation{Universidad Aut\'{o}noma del Estado de Hidalgo, Pachuca, M\'{e}xico }
\email{ropapaqui@gmail.com}

\author[0000-0002-9105-0518]{L.~Olivera-Nieto}
\affiliation{Max-Planck Institute for Nuclear Physics, 69117 Heidelberg, Germany}
\email{laura.olivera-nieto@mpi-hd.mpg.de}

\author[0000-0002-5448-7577]{N.~Omodei} 
\affiliation{Department of Physics, Stanford University: Stanford, CA 94305–4060, USA}
\email{nicola.omodei@stanford.edu}

\author[0009-0009-2481-6921]{M.~Osorio}
\affiliation{Instituto de Astronom\'{i}a, Universidad Nacional Autónoma de México, Ciudad de Mexico, Mexico}
\email{jmosorio@astro.unam.mx}

\author[]{E.~Ponce}
\affiliation{Facultad de Ciencias F\'{i}sico Matemáticas, Benemérita Universidad Autónoma de Puebla, Puebla, Mexico}
\email{eponce@fcfm.buap.mx}

\author[0000-0002-8774-8147]{Y.~Pérez Araujo}
\affiliation{Instituto de F\'{i}sica, Universidad Nacional Autónoma de México, Ciudad de Mexico, Mexico}
\email{yuniorpy@gmail.com}

\author[0000-0001-5998-4938]{E.G.~Pérez-Pérez} 
\affiliation{Universidad Politecnica de Pachuca, Pachuca, Hgo, Mexico}
\email{egperezp@yahoo.com.mx}

\author[0000-0002-6524-9769]{C.D.~Rho}
\affiliation{Department of Physics, Sungkyunkwan University, Suwon, Gyeonggi, South Korea}
\email{no397@naver.com}

\author[0000-0003-1327-0838]{D.~Rosa-Gonz\'{a}lez}
\affiliation{Instituto Nacional de Astrof\'{i}sica, \'{O}ptica y Electr\'{o}nica, Puebla, M\'{e}xico }
\email{danrosa@inaoep.mx}

\author[0000-0002-4204-5026]{M.~Roth} 
\affiliation{Los Alamos National Laboratory, Los Alamos, NM, USA}
\email{mattroth@lanl.gov}

\author[0000-0001-6939-7825]{E.~Ruiz-Velasco}
\affiliation{Max-Planck Institute for Nuclear Physics, 69117 Heidelberg, Germany}
\email{edna.ruiz@mpi-hd.mpg.de}

\author{H.~Salazar}
\affiliation{Facultad de Ciencias F\'{i}sico Matem\'{a}ticas, Benem\'{e}rita Universidad Aut\'{o}noma de Puebla, Puebla, M\'{e}xico }
\email{hsalazar@fcfm.buap.mx}

\author[0000-0001-6079-2722]{A.~Sandoval}
\affiliation{Instituto de F\'{i}sica, Universidad Nacional Aut\'{o}noma de M\'{e}xico, Ciudad de M\'{e}xico, M\'{e}xico }
\email{asandoval@fisica.unam.mx}

\author[0000-0001-8644-4734]{M.~Schneider}
\affiliation{Dept. of Physics, University of Maryland, College Park, MD 20742, USA}
\email{mschnei4@umd.edu}

\author{G.~Schwefer} 
\affiliation{Max-Planck Institute for Nuclear Physics, 69117 Heidelberg, Germany}
\email{georg.schwefer@mpi-hd.mpg.de}

\author{J.~Serna-Franco} 
\affiliation{Instituto de F\'{i}sica, Universidad Nacional Autónoma de México, Ciudad de Mexico, Mexico}
\email{j_serna@ciencias.unam.mx}

\author[0000-0002-1012-0431]{A.J.~Smith}
\affiliation{Dept. of Physics, University of Maryland, College Park, MD 20742, USA}
\email{asmith8@umd.edu}

\author{Y.~Son}
\affiliation{Natural Science Research Institute, University of Seoul, Seoul, Republic of Korea}
\email{youngwan.son@cern.ch}

\author[0000-0002-1492-0380]{R.W.~Springer}
\affiliation{Department of Physics and Astronomy, University of Utah, Salt Lake City, UT, USA }
\email{wayne.springer@utah.edu}

\author{O.~Tibolla}
\affiliation{Universidad Politecnica de Pachuca, Pachuca, Mexico}
\email{omar.tibolla@gmail.com}

\author[0000-0001-9725-1479]{K.~Tollefson}
\affiliation{Dept. of Physics and Astronomy, Michigan State University, East Lansing, MI 48824, USA}
\email{tollefson@pa.msu.edu}

\author[0000-0002-1689-3945]{I.~Torres}
\affiliation{Instituto Nacional de Astrof\'{i}sica, \'{O}ptica y Electr\'{o}nica, Puebla, M\'{e}xico }
\email{ibrahim.torres23@gmail.com}

\author[0000-0002-7102-3352]{R.~Torres-Escobedo}
\affiliation{Tsung-Dao Lee Institute \& School of Physics and Astronomy, Shanghai Jiao Tong University, 800 Dongchuan Rd, Shanghai, SH 200240, China }
\email{torresramiro350@sjtu.edu.cn}

\author[0000-0003-1068-6707]{R.~Turner}
\affiliation{Department of Physics, Michigan Technological University, Houghton, MI, USA }
\email{rturner1@mtu.edu}

\author[0000-0001-6798-353X]{X.~Wang}
\affiliation{Department of Physics, Michigan Technological University, Houghton, MI, USA }
\email{xwang32@mtu.edu}

\author{Z.~Wang}
\affiliation{Department of Physics, University of Maryland, College Park, MD, USA}
\email{zhen@umd.edu}

\author[0000-0003-2141-3413]{I.J.~Watson} 
\affiliation{University of Seoul, Seoul, Rep. of Korea}
\email{ian.james.watson@cern.ch}

\author[0009-0005-7243-1402]{H.~Wu} 
\affiliation{Department of Physics, University of Wisconsin-Madison, Madison, WI 53706, USA }
\email{hwu298@wisc.edu}

\author[0009-0006-3520-3993]{S.~Yu} 
\affiliation{Department of Physics, Pennsylvania State University, University Park, PA, USA}
\email{sjy5345@psu.edu}

\author[0000-0002-9307-0133]{S.~Yun-C\'{a}rcamo} 
\affiliation{Department of Physics, University of Maryland, College Park, MD, USA}
\email{yunsoh@umd.edu}

\author[0000-0003-0513-3841]{H.~Zhou}
\affiliation{Tsung-Dao Lee Institute \& School of Physics and Astronomy, Shanghai Jiao Tong University, Shanghai, China}
\email{hao_zhou@sjtu.edu.cn}

\author[0000-0002-8528-9573]{C.~de Le\'{o}n}
\affiliation{Universidad Michoacana de San Nicol\'{a}s de Hidalgo, Morelia, M\'{e}xico } 
\email{cederik.de.leon@umich.mx}

\collaboration{all}{HAWC Collaboration}

\begin{abstract}

Supernova remnants are one potential source class considered a PeVatron (i.e. capable of accelerating cosmic rays above PeV energies). The shock fronts produced after the explosion of the supernova are ideal regions for particle acceleration. IC 443 is a supernova remnant that has been studied extensively at different wavelengths. Using 2966 days of gamma-ray data from the HAWC observatory, we study the emission of IC 443 with the objective of finding signatures of cosmic-ray acceleration at the PeV scale. Using a maximum likelihood method, we find a point source located at ($\alpha$=94.42$^{\circ}$, $\delta$=22.35$^{\circ}$) that we associate with IC 443. The measured spectrum is a simple power law with an index of $-3.14\pm$0.18, which is consistent with previous TeV observations.  Although we cannot confirm that IC 443 is a hadronic PeVatron, we do not find any sign that the spectrum has a cut off at tens of TeV energies, with the spectrum extending to $\sim$30 TeV. Furthermore, we also find a new extended component in the region whose emission is described by a  simple power law with an index of $-2.49\pm$0.08 and which we call HAWC J0615+2213. While we show evidence that this new source might be a new TeV halo, we defer a detailed analysis of this new source to another publication.

\end{abstract}



\section{Introduction}\label{sec:intro}

A supernova explosion occurs when the gravitational force of a star overcomes the internal pressure supporting it against collapse. This loss of pressure happens when a massive star exhausts its thermonuclear fuel. Alternatively, a gravitational collapse may happen when a white dwarf accretes enough mass from a companion star, surpassing its internal pressure. The aftermath of this explosion produces a supernova remnant, composed of ejected material expanding outward and bounded by a shock wave, along with the shocked gas that once surrounded the star.  Supernova remnants are considered one of the primary sources of galactic cosmic rays and are one of the candidates to be PeVatrons---astrophysical sources capable of accelerating cosmic rays up to PeV energies---depending on the type of the supernova~\citep{lowRatePeV,GalPeVatrons}.

IC 443---the Jellyfish Nebula---is a supernova remnant discovered by the German-American astronomer Max Wolf in 1892 \citep{maxWolfIC433}. The position of IC 443, as it appears in the Green Catalog of Supernova Remnants~\citep{snrGreen}, is at ($\alpha$=94.25$^{\circ}$, $\delta$=22.56$^{\circ}$) in celestial coordinates (J2000). This remnant has been studied in detail across the entire electromagnetic spectrum. 

Radio observations have mapped the cold gas and dust structures within the supernova remnant. The radio emissions indicate regions of synchrotron radiation, where relativistic electrons spiral around magnetic fields, and provide information about the magnetic environment of the remnant~\citep{radioClark,radioDat,radioIC443Cont}. 
Optical observations have helped delineate the boundaries of the supernova remnant and identify regions of high-energy interactions between the surrounding interstellar medium and the shock waves~\citep{opticalIC443}. 
Infrared observations highlight the presence of warm dust and molecular gas within the supernova remnant. They also reveal regions where shock waves from the explosion heat the surrounding material and show the distribution of cold, dense molecular clouds interacting with the expanding shock fronts~\citep{irData}.
X-ray observations of IC 443, such as those by \cite{xrayIc443I,nustar}, and \cite{xmmnewton}, reveal the hot, high-energy gas created by the shock waves of the supernova explosion. Furthermore, these observations have helped identify the progenitor of IC 443 as the neutron star CXOU J061705.3+222127~\citep{chandraPulsar}. They have also provided additional evidence that IC 443 was a Type II supernova~\citep{ic443TypeII}.
High-energy observations of this supernova remnant were performed by VERITAS~\citep{veritasIC443}, MAGIC~\citep{magicIC443}, and Fermi-LAT~\citep{fermiIC443}. Gamma-ray emission in the Fermi energy band was confirmed to be the result from the interaction of hadronic cosmic rays accelerated in the SNR with molecular clouds surrounding the remnant after observing the pionic bump in the spectral energy distribution~\citep{pionSNR}.

Other studies have shown the complexity in this region, with the supernova remnant G189.6+03.3 overlapping with IC 443 and possibly interacting with it \citep{xrayG189IC443}. These studies suggest that the distances to both objects are comparable, with distances between 1.5~kpc to 2.5~kpc \citep{ic443optKinDist}. Their ages are uncertain, ranging from a minimum of 3 kyr and a maximum of about 30~kyr \citep{xrayG189IC443}. \cite{modelMixedXRay} measured an age of $\sim$8~kyr after comparing a hydrodynamic model for IC 443 describing the interaction of the remnant with the environment and comparing the expected X-ray emission with XMM-Newton data.

In this work, we report our observations of this region using HAWC data. 
Unexpectedly, we have also found a new gamma-ray component with an extended morphology and a hard spectral flux. We investigate three hypotheses to explain this emission: (1) secondary gamma rays from cosmic rays interacting with gas in the region; (2) a population of faint sources ;(3) a new TeV halo produced by the pulsar B0611+22. This pulsar was discovered by \cite{b0611P22}. It is $\sim$90~kyr old and is located at 3.55~kpc \citep{pulsarDist}\footnote{The pulsar has a dispersion measure of 96.91 $\rm pc \, cm^{-3}$, which puts the pulsar at 1.74~kpc or 3.5~kpc, depending on the model used to measured the distance \citep[see][]{electronDens}. However, \cite{pulsarDist} have parallax measurements and hence we decided to use this value.}. Most recent observations on these pulsar have been in radio and X-rays, with the latter only showing upper limits \citep{psrb0611radioxrays}.

%__________________________________________________________________
\section{The HAWC Observatory and the Dataset}\label{sec:hawc}
The High Altitude Water Cherenkov (HAWC) Observatory is a gamma-ray detector located at 4100~m a.s.l. in central Mexico, near the Pico de Orizaba volcano. The detector consists of 300 steel tanks containing 180,000~litres of water, which serves as the refractive medium used to produce Cherenkov light. This light is detected by four photomultipliers (PMTs), three of which are eight inches and one being ten inches in size. The PMTs are anchored at the bottom of the tanks and facing upwards.  When a gamma-ray interacts with the atmosphere, it produces a cascade of particles, known as an extensive air shower. The secondary particles propagate and spread out through the atmosphere, expanding the shower front up to a point where the energy is no longer sufficient. The particles in the shower front penetrate the steel tanks and produce Cherenkov light in the water that the PMTs detect. The detector is sensitive to gamma rays with energies between 300~GeV to $>$100~TeV. It has a high-duty cycle of about $>$95\% and an instantaneous field of view of 2~sr, covering 2/3 of the sky every day \citep{hawc,3hwc,HAWCNIM}.
 
We recover the information of the primary gamma ray---its location in the sky and energy---by using the triggered time and the measured charge in each PMT. The energy of the gamma ray is reconstructed with two independent methods. One is by fitting the shower profile after locating the core of the shower in the array. The other is a neural network to estimate the energy of the gamma ray. Both methods are described in \cite{hawcp5}.

For the analysis, we have used two binning schemes. The first binning scheme is based on the ratio between triggered and total active PMTs during a shower event. We denote this ratio as $f_{\rm hit}$. In this scheme, we consider events whose reconstructed cores are measured inside and outside the main array.  For the second binning scheme, along with the $f_{\rm hit}$ information, we also bin the events according to their reconstructed energy, creating what we refer to as energy scheme. This second scheme only considers events with reconstructed cores inside the main array.
We use both schemes for our analysis. The first scheme is optimized to detect 2\% of the Crab Nebula flux, and hence we use it to find the best-fit morphology of the sources in the region. The second scheme is better suited to obtain spectral points since we have more accurate energy information. All the schemes give a similar flux measurement as seen in Figure 9 of the HAWC performance paper \citep{hawcp5}.

%
%_______________________________________________________________
%_______________________________________________________________
\section{Analysis and Results}\label{sec:results}
\subsection{Analysis Method}
We use data collected from 2015 to 2023. The livetime is 2966 days using the $f_{\rm hit}$ bin scheme, while with the reconstructed energy scheme, it is 2769 days since we remove data from the time period where HAWC was under construction and not all tanks were operational. 

We apply a gamma-hadron selection cut found in \cite{hawcp5}, and estimate the background using the method described in section 4.3 of \citep{crab100tev}. 
Once we have the estimated background, we proceed with the analysis based on the maximum likelihood estimation method. The analysis is performed using the Multi-Mission Maximum Likelihood (threeML) framework with the HAWC Accelerated Likelihood (HAL plugin) as described in \cite{threeml,hal}. 
ThreeML uses the likelihood ratio test to determine whether one model is better than another. The test statistic is defined as 
\begin{equation}\label{eq:TS}
TS=2(\ln L_{\rm i} - \ln L_{\rm j}), 
\end{equation}
where $L_{i,j}$ are the maximum likelihoods of models $i, j$. It is important to remark that we also use this statistic when generating sky maps, where the calculation is performed for each individual pixel.

The algorithm we use to determine the morphology of the region is described in \cite{hawcanalysis}, with the difference that we now use the High-Energy Radiative Messengers (HERMES) diffuse emission template as the galactic diffuse emission (GDE) model \citep{hermes}.  HERMES is a spatial propagation template based on the DRAGON cosmic-ray propagation code~\citep{dragon}. The model includes pionic gamma rays, produced by the interaction of the sea of cosmic rays with neutral and molecular hydrogen. It also includes inverse Compton from the cosmic-ray leptons scattering on low-energy photons in the Galaxy such as UV, optical, infrared or microwave photons.  This new GDE model has the advantage that it considers the morphology of the gas in the region, whereas \cite{hawcanalysis} used a Gaussian shape as a function of galactic latitude. When fitting the GDE model, we free a unit-less factor parameter that scales the normalization of the baseline diffuse flux calculated by HERMES.

To summarize the algorithm used in \cite{hawcanalysis}, we first search for all possible point sources in the region and assume a simple power-law spectrum for each one as shown in Algorithm \ref{alg:ps}. Then, each point source is tested as an extended source as described in Algorithm \ref{alg:es}. Once we have found the final morphology of each source, we test for curvature in the spectrum, with the process described in Algorithm \ref{alg:spectrum}.
\begin{algorithm}[!htbp]
\caption{Point Source Morphology Search}\label{alg:ps}
\KwData{Initial model with fitted GDE model}
\KwResult{Optimized model with point sources, if present}
\While{TS$_{\rm model}$ $>$ 25}{
    Add a point source with a simple power-law spectrum at the hotspot with maximum significance in the skymap\;
    
    The spectrum of the added source is modeled as Eq.~\ref{eq:spl}\;
    
    Calculate TS$_{\rm model}$ between model with extra point source and previous accepted model\;
    \If{TS$_{\rm model}>$25} {Keep new model\;}
}
Move to extended source morphology search\;
\end{algorithm}

The simple power-law spectrum is defined as
\begin{equation}\label{eq:spl}
\Phi_{SPL} = \Phi_0 \left( \frac{E}{E_{\rm piv}} \right)^{\alpha}
\quad \text{[TeV$^{-1}$ cm$^{-2}$ s$^{-1}$]},
\end{equation}
where $\Phi_0$ is the normalization at the pivot energy $E_{\rm piv}$, and $\alpha$ is the spectral index.

\begin{algorithm}[!htbp]
\caption{Extended Source Morphology Search}\label{alg:es}
\KwData{Initial model with fitted point sources}
\KwResult{Optimized model with extended sources, if present}

Identify the point source with the highest TS and replace it with a Gaussian spatial morphology (see Appendix for the functional form), keeping all other sources as point-like. Fix all source positions and fit the spectra and extension parameters\;

Compute the TS between the previous model and the new model\;
\If{TS$_{\rm model}$ $>$ 16}{
    Accept the new model\;
}
\Else{
    Reject the extended source model and proceed to the point source with the next highest TS$_{\rm source}$\;
    Return to Step 1\;
}
Evaluate the TS$_{\rm source}$ of all remaining point sources in the accepted model\;
\If{TS$_{\rm source}$ $>$ 16 for all point sources}{
    Return to Step 1\;
}
\ElseIf{TS$_{\rm source}$ $<$ 16 for any point source}{
    Remove the low-TS source(s) and refit the model, floating all parameters\;
    \If{point sources remain}{
        Return to Step 10\;
    }
    \Else{
        Terminate the study as no point sources remain for extension testing\;
    }
}
\end{algorithm}

TS$_{\rm source}$ is calculated by comparing the likelihood values of the source in question when it is present in the model versus when is not in the model.

The final step is to check whether the spectra of the sources has a cutoff or curvature. We have tested for a logarithmic parabola and a power law with exponential cutoff, whose functional forms are:
\begin{equation}
	\Phi = \Phi_{SPL} \left( \frac{E}{E_{\rm piv}}\right)^{\beta\ln(E/E_{\rm piv})},
\end{equation}
where $\Phi_{SPL}$ is the same as in Equation \ref{eq:spl}
and $\beta$ is the curvature parameter;
\begin{equation}
	\Phi = \Phi_{SPL}\exp \left(-\frac{E}{E_{\rm cutoff}}\right),
\end{equation}
where $\Phi_{SPL}$ is the same as in Equation \ref{eq:spl} and $E_{\rm cutoff}$ is the cutoff energy.

\begin{algorithm}[!h]
\caption{Spectrum Test}\label{alg:spectrum}
\KwData{Best-morphology model found}
\KwResult{Optimized model with curvature in the spectrum if observed}
Fix the positions of all sources\;

Calculate the log-likelihood of the model assuming all sources have simple power-law spectra\;

Select the source with the highest TS$_{\rm source}$\;

Change the spectral model of this source from a simple power law to a model with curvature, and refit the model\;

\While{TS$_{\rm model}$ $>$ 16 between the new model and the old model}{
    Keep the new model\;
    Select the source with the highest TS$_{\rm source}$ among the remaining sources\;
    Change its spectrum from a simple power law to a curved model, and refit\;
}
\end{algorithm}

\subsection{Observations}

\begin{figure*}[!htbp]
	\centering
	\plottwo{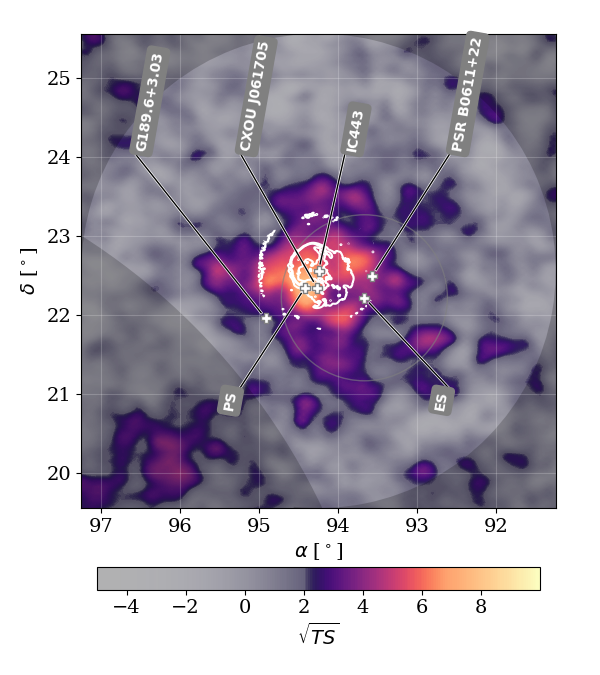}{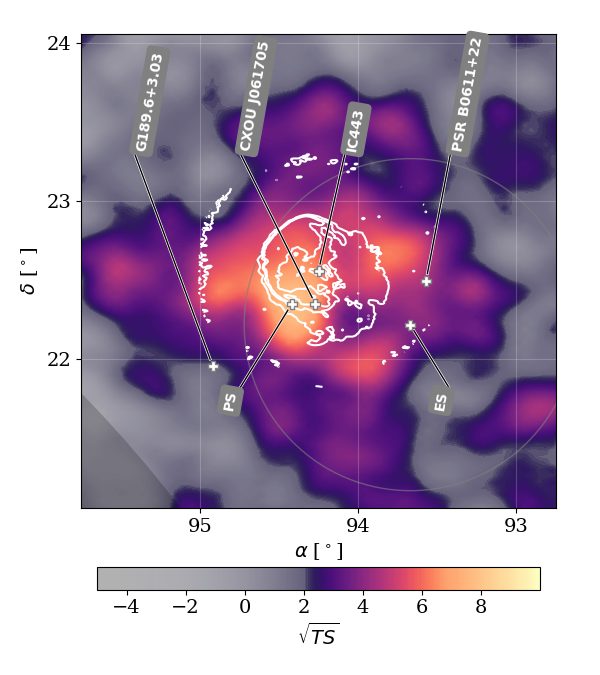}
	\caption{IC 443 region seen by HAWC using the $f_{\rm hit}$ analysis bins. The map assumes a point source morphology with an index of -3.14. The radio contours (in white) are from \cite{radioIC443Cont}. Left: The region of interest used for the analysis is also shown with the change of brightness in the sky map. The positions of IC 443 and G189.6+03.3 are from the Green Catalog \cite{snrGreen}. Position of CXOU J061705.3+222122 is from \cite{chandraPulsar}. Position of the pulsar B0611+22 comes from \cite{b0611P22}. The positions found in the analysis for the point source (PS) and extended source (ES) are also shown. Gray circle is the $\sigma$ width of the ES. Right: a zoom in on the region centered at the position of IC 443.}\label{fig:ic443}
\end{figure*}

Figure \ref{fig:ic443} shows the HAWC $f_{\rm hit}$ significance map of the region of IC 443. The highlighted area indicates the region of interest used for the analysis. 
The best-fit model describing the region together with the GDE model is a point source (PS) and an extended source (ES). 
We found no significant preference for a spectral curvature for either source. 
We associate the point source with IC 443. For the extended source, which we name HAWC J0615+2213, we did not find a possible counterpart in the literature. Therefore, we discuss three possibilities: emission originating from the interaction of freshly injected cosmic rays with the interstellar medium, unresolved sources, or a new TeV Halo object. We defer this discussion to Sect. \ref{sec:discussion}.

The position of the point source is located 0.26$^{\circ}$ away from the Green catalog position of IC 443 and 0.14$^{\circ}$ from the pulsar wind nebula associated with the neutron star CXOU J061705.3+222212. The centroid of the extended source is 0.63$^{\circ}$ from IC 443, 0.57$^{\circ}$ from CXOU J061705.3+222212, and 0.29$^{\circ}$ from the pulsar B0611+22. The source parameter results are summarized in Table \ref{tab:results1} together with the TS of each source. The pivot energy, $E_{\rm piv}$, that reduces the correlation between the parameters was found to be 2.3 TeV.
Table \ref{tab:resultsSpectra} shows the comparison of the fitted spectral parameters between the $f_{\rm hit}$ binning scheme and the energy binning scheme. 

\begin{figure*}[!htbp]
\centering
	\includegraphics[scale=0.45]{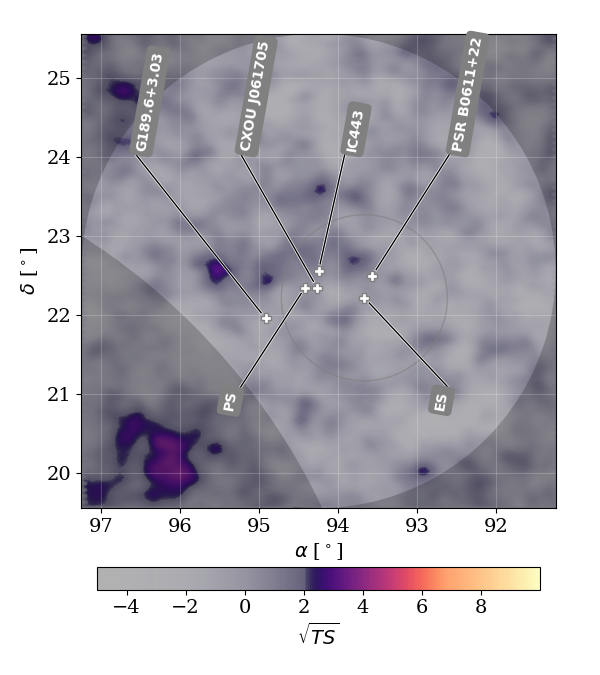}
	\includegraphics[scale=0.45]{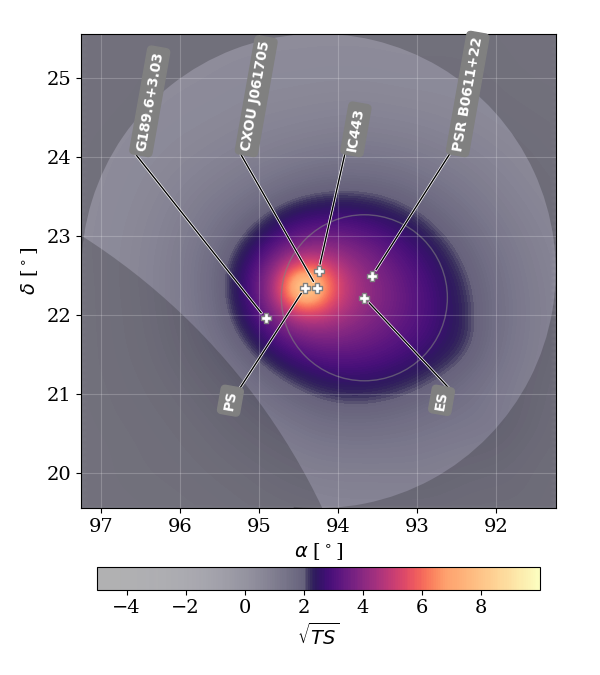}
	\caption{Left: Residual sky map after fitting the best-fit model using the $f_{\rm hit}$ analysis bins. The region of interest used for the analysis is also shown with the change of brightness in the skymap. Right: Best-fit sky map model. 
	%The histogram is the significance distribution of the region. 
	No significant excesses are observed after subtracting the best-fit model.}\label{fig:residual}
\end{figure*}

\begin{figure*}[!htbp]
\centering
    \plottwo{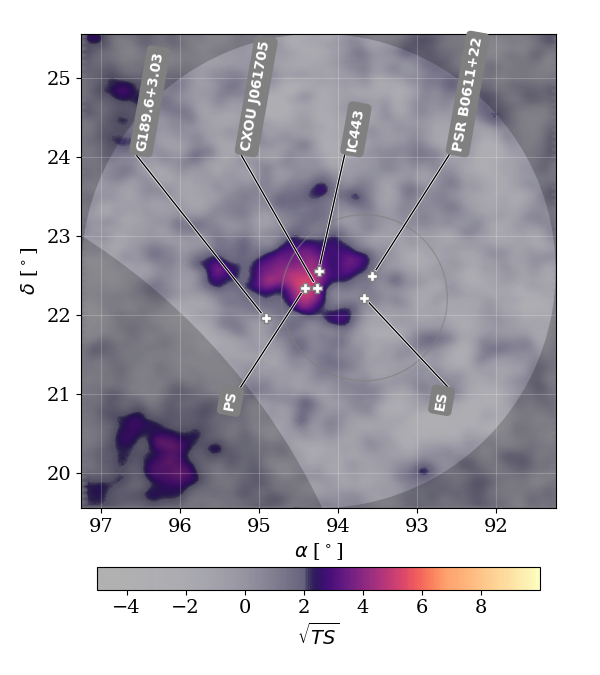}{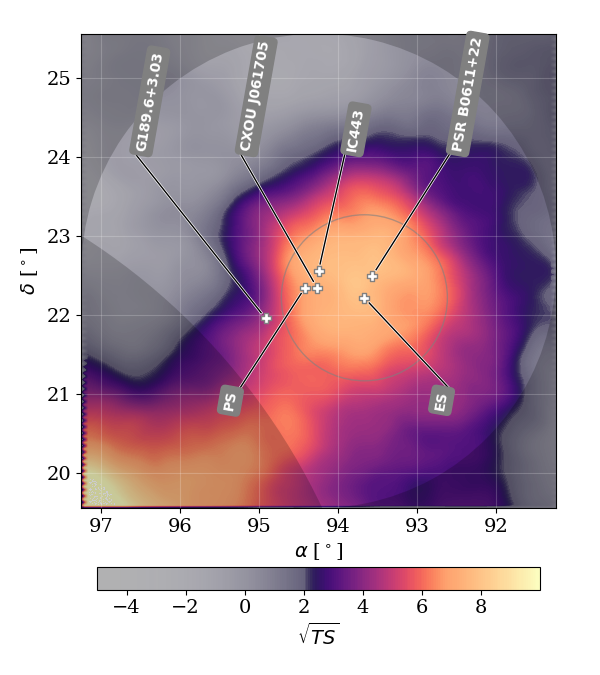}
	\caption{Residual maps of the region for the fhit analysis bins. Left: subtracting the extended source. Right: subtracting the point source. In the second significance map, we made it assuming an extension of 1.05$^{\circ}$. On the lower left of the plot the Geminga Halo is visible. The region of interest used for the analysis is shown with the change of brightness in the sky map.}\label{fig:residualIndSources}
\end{figure*}

Figure~\ref{fig:residual} shows the map of the best-fit model, as well as the residual map of the region, and the significance distribution of the sky. The significance distribution is relatively close to a Gaussian distribution, showing that the model adequately describes most of the signal and the remaining is only background. Figure~\ref{fig:residualIndSources} shows each source after subtracting either the extended source or the point source from the original map.

\begin{deluxetable*}{ccccccc}[!htbp]
\tablewidth{0pt}
\tablecaption{Results of the analysis in the region of IC 443 using the $f_{\rm hit}$ scheme.}
\tablehead{
\colhead{Source} & \colhead{$\Phi_0$} & \colhead{Index} & \colhead{R.A.} & \colhead{Decl.} & \colhead{$\sigma$} & TS$_{\rm source}$\\
\colhead{} & \colhead{[$\rm TeV^{-1} \, cm^{-2} \, s^{-1}$]} & \colhead{} & \colhead{[$^\circ$]} & \colhead{[$^\circ$]} & \colhead{[$^\circ$]}
}
\startdata
PS & $(5.9 \pm 1.3 \,{^{+0.35}_{-0.91}}) \times 10^{-14}$ & $-3.14 \pm 0.18 \,{^{+0.08}_{-0.09}}$ & $94.42{^{+0.07}_{-0.05}} \,{\pm 0.01}$ & $22.35{^{+0.06}_{-0.07}} \,{^{+0.05}_{-0.03}}$ & -- & 28.2 \\
ES & $(3.18^{+1.37}_{-0.92} \,\pm 1.3) \times 10^{-13}$ & $-2.49 \pm 0.08 \,^{+0.01}_{-0.03}$ & $93.67 \pm 0.19 \,\pm 0.04$ & $22.22 \pm 0.20 \,\pm 0.1$ & $1.05^{+0.21}_{-0.18} \,\pm 0.18$ & 88.8 \\
GDE model & $2.62 \pm 1.20 \,{^{+0.07}_{-0.30}}$* & -- & -- & -- & -- & 16.5\\
\enddata
\tablecomments{First set of uncertainties are statistical while the second set are systematic. The normalization of the two sources is at $E_{\rm piv} = 2.3$ TeV. PS is associated with IC 443. ES is the new source HAWC J0615+2213.\\
*Scale factor: unitless and only includes systematics from detector configuration.}
\label{tab:results1}
\end{deluxetable*}

\begin{deluxetable*}{ccc}[!htbp]
\tablewidth{0pt}
\tablecaption{Comparison of the spectral parameter results between each binning scheme.}
\tablehead{
\colhead{Parameter} & \colhead{$f_{\rm hit}$ Scheme} & \colhead{Energy Scheme}
}
\startdata
PS $\Phi_0$ & $(5.9 \pm 1.3) \times 10^{-14}$ & $(7.8^{+1.5}_{-1.3}) \times 10^{-14}$ \\
PS Index & $-3.14 \pm 0.18$ & $-3.07 \pm 0.12$ \\
ES $\Phi_0$ & $(3.18^{+1.37}_{-0.92}) \times 10^{-13}$ & $(3.0^{+0.8}_{-0.6}) \times 10^{-13}$ \\
ES Index & $-2.49 \pm 0.08$ & $-2.61 \pm 0.09$ \\
\enddata
\tablecomments{Comparison between the two analysis binning schemes. Uncertainties are statistical only. Units of $\Phi_0$ are in $\rm [TeV^{-1} \, cm^{-2} \, s^{-1}]$.}
\label{tab:resultsSpectra}
\end{deluxetable*}

The energy ranges in which the sources are significantly detected are calculated using the method described in \cite{hawc}. At the 1$\sigma$, or 68\% confidence level, these are [0.3,30] TeV for the point source and [0.7,47] TeV for the extended source. 

%\textbf{We observe no significant excess in the direction of the supernova remnant G189.6+03.3. We calculate 95\% upper limits which we present in Sec. \ref{sec:g189p6}.}
We observe no significant excess in the direction of the supernova remnant G189.6+03.3. We calculate 95\% upper limits which we present in Sec. \ref{sec:g189p6}.

\subsubsection{Systematics}\label{sec:results.syst}

To account for detector biases, we performed the analysis with multiple detector responses that consider different calibration, detector and simulation effects. This includes calibration performance, charge uncertainty, PMT efficiency and PMT threshold. 

We also estimated a systematic based on the modeling of the region by setting the scaling factor of the GDE model to one and by fitting individually the different components of the GDE model which include inverse Compton, pion decay from atomic hydrogen and pion decay from molecular hydrogen. We find the difference between the default and systematic results and add them in quadrature to determine the final systematic uncertainties~\citep[see][for more information]{hawc,3hwc}. The uncertainties are listed in Table \ref{tab:results1}.

\subsection{Spectral Energy Distributions (SED)}
\begin{figure*}[!htbp]
\centering
    \plottwo{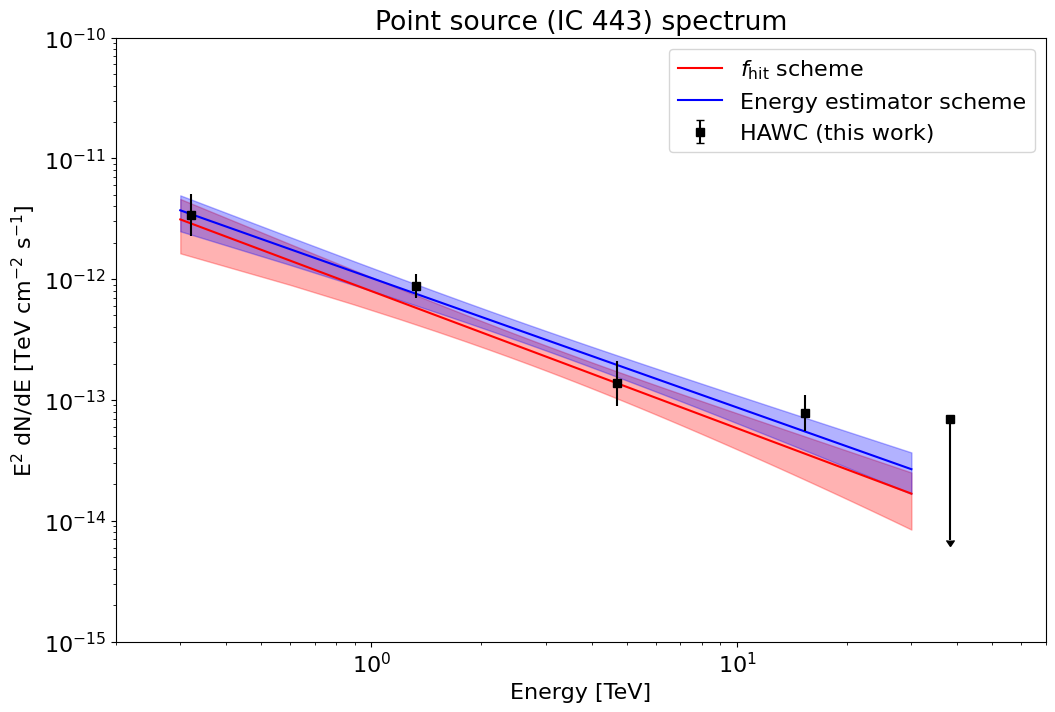}{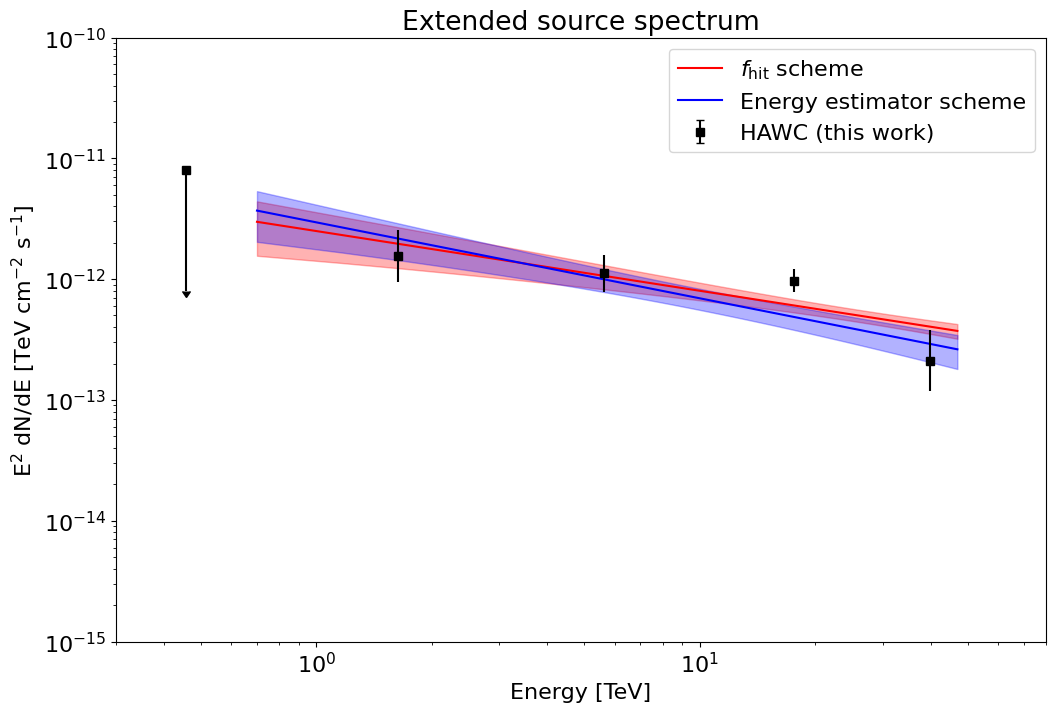}
	\caption{Spectrum of the two sources seen by HAWC. Left is the spectrum of the point source, which we associate with IC 443. Right is the spectrum of the extended source.  }\label{fig:spectra}
\end{figure*}

As mentioned before, the 1D binning scheme is used to obtain the best spectral shape of a source with threeML. To obtain the spectral points, we use the 2D binning scheme. We fix all the parameters of the model except for the normalization of the source in each energy bin \citep[See][for a description of this algorithm]{crab100tev}. 
Tables \ref{tab:results2} and \ref{tab:results3} present the differential particle flux measured by HAWC and the E$^2$-weighted differential particle flux is shown in Fig.~\ref{fig:spectra}. The figure also compares the spectral model fitted with the $f_{\rm hit}$ scheme and the energy bin scheme, showing that they are consistent with each other. 

\begin{deluxetable}{ccc}[!htbp]
\tablewidth{0pt}
\tablecaption{SED flux points of the PS IC 443\label{tab:results2}}
\tablehead{
\colhead{Energy} & \colhead{Flux} & \colhead{TS} \\
\colhead{[TeV]} & \colhead{[TeV$^{-1}$ cm$^{-2}$ s$^{-1}$]} & \colhead{}
}
\startdata
0.3 & $(3.3^{+1.6}_{-1.1})\times10^{-11}$ & 6.2 \\
1.3 & $(5.0^{+1.3}_{-1.0})\times10^{-13}$ & 18.5 \\
4.7 & $(6.2^{+3.3}_{-2.1})\times10^{-15}$ & 5.4 \\
15.4 & $(3.3^{+1.4}_{-1.0})\times10^{-16}$ & 11.1 \\
38.2 & $<4.5\times10^{-17}$ & 0.3 \\
\enddata
\tablecomments{Uncertainties are statistical only. Upper limits (U.L.) are given at 90\% confidence level (C.L.).}
\end{deluxetable}

%\vspace{1em}

\begin{deluxetable}{ccc}[!htbp]
\tablewidth{0pt}
\tablecaption{SED flux points of the ES HAWC J0615+2213\label{tab:results3}}
\tablehead{
\colhead{Energy} & \colhead{Flux} & \colhead{TS} \\
\colhead{[TeV]} & \colhead{[TeV$^{-1}$ cm$^{-2}$ s$^{-1}$]} & \colhead{}
}
\startdata
0.5 & $<3.8\times10^{-11}$ & 0.5 \\
1.6 & $(5.8^{+3.8}_{-2.3})\times10^{-13}$ & 3.8 \\
5.6 & $(3.5^{+1.5}_{-1.0})\times10^{-14}$ & 8.4 \\
17.6 & $(3.1^{+0.8}_{-0.6})\times10^{-15}$ & 23.1 \\
39.8 & $(1.4^{+1.0}_{-0.6})\times10^{-16}$ & 3.2 \\
\enddata
\tablecomments{Uncertainties are statistical only. Upper limits (U.L.) are given at 90\% confidence level (C.L.).}
\end{deluxetable}

%
%_______________________________________________________________
\section{Discussion}\label{sec:discussion}
\subsection{Multi-wavelength observations of IC 443}

Several researchers have performed extensive multi-wavelength studies of IC 443 over the years. Early works by \cite{snrMC} and \cite{snrModeling} laid the groundwork of multi-wavelength observations, focusing mainly on the radio and X-ray components; while more recent studies as \cite{pionSNR, nustar} and \cite{MWGSNR}, have furthered our understanding at higher energies up to 1 TeV. We use the new HAWC observations to extend the multi-wavelength analysis up to $30$ TeV. As mentioned earlier, we associate the point source we found in Sect. \ref{sec:results} with IC 443.

In the very-high-energy ($>$100 MeV) part of the electromagnetic spectrum, \cite{pionSNR} demonstrated that gamma-ray emission is primarily due to the decay of neutral pions produced in inelastic collisions between cosmic-ray protons and dense molecular clouds in the region \citep{molCloudI, modelMixedXRay}. They also explored leptonic models, showing that these models could not account for the observed gamma-ray flux due to energetic constraints \citep[see][]{W51c}. These constraints arise from the maximum energy that electrons can achieve in an environment like IC 443. Electron acceleration is restricted by the underlying mechanism, such as diffusive shock acceleration (DSA), and by substantial energy losses due to synchrotron radiation, inverse Compton scattering, and bremsstrahlung. These losses are further amplified by the dense molecular cloud surrounding IC 443.

Under typical magnetic fields ($\sim$100 $\mu$G) and accounting for both synchrotron and inverse Compton losses, the maximum energy an electron can gain is approximately 5 TeV \citep{Petrosian2012, Yamaguchi2009}. When bremsstrahlung losses are included in regions with higher gas densities (up to 100 cm$^{-3}$) \citep{Troja2008}, this maximum decreases further to around 3 TeV. Even with non-linear shock effects that amplify the magnetic field up to ~300 $\mu$G, the maximum achievable electron energy is only $\sim$10 TeV. Thus, while electrons can contribute to high-energy emission, they cannot explain gamma rays up to 30 TeV. Furthermore, according to \cite{Yamaguchi2009}, the measured magnetic field strength in the region of IC 443 is 10$\mu$G, constraining even more the maximum energy of the electrons. This limitation makes hadronic processes a more plausible explanation for the observed gamma-ray spectrum \citep{Guo2014, Zhang2011}.

Protons can achieve significantly higher energies due to their relatively low energy losses. The maximum energy for protons, achievable via DSA, depends on factors such as shock speed, magnetic field strength, and remnant size. For IC 443, typical values include a shock speed of 300 km s$^{-1}$ \citep{Lee2008, Troja2008}, a magnetic field strength of 10 $\mu$G \citep{Yamaguchi2009}, and a remnant size of 10.9 pc \citep{Troja2008}. These parameters suggest that protons could theoretically reach energies up to $\sim$65 TeV \citep{Drury1983, Lagage1983, Bell2013}. With the inclusion of non-linear diffusive shock acceleration (NLDSA) effects, which amplify the magnetic field, this maximum energy may extend to 100 TeV or even higher, potentially approaching 1 PeV \citep{Bell2004, Bell2013, Caprioli2009}.

To test these estimates of the maximum proton energies, we perform an analysis using the Markov Chain Monte Carlo (MCMC) software \texttt{Naima} \citep{naima}. This software fits data to non-thermal emission models by optimizing the parameters of the parent particle spectra. Additionally, it calculates the Bayesian Information Criterion (BIC) for the models, which we use to compare and evaluate their performance.

We incorporate observations from Fermi \citep{fermiIC443}, VERITAS \citep{veritasIC443}, MAGIC \citep{magicIC443}, and the results presented here. As a side note, the recently commissioned LHAASO detector---a gamma-ray observatory with better sensitivity above 10 TeV compared to HAWC---has reported extended emission in the region as part of its catalog search. LHAASO observes a source with a width of $(0.59\pm0.08)^{\circ}$ located at coordinates 94.35$^{\circ}$, 22.57$^{\circ}$. Their analysis using LHAASO-WCDA data, based on a simple power-law model, yielded a differential particle flux at 3 TeV of (1.95$\pm$0.27)$\times10^{-13}$ TeV$^{-1}$ cm$^{-2}$ s$^{-1}$ with a spectral index of $-2.92\pm0.14$ in the energy range between 1~TeV to 25~TeV. An upper limit of 0.17~$\times10^{-15}$ TeV$^{-1}$ cm$^{-2}$ s$^{-1}$ at 50 TeV was set using LHAASO-KM2A~\citep{lhaasoCat}. However, further detailed analysis is required to determine whether this observation aligns with the morphology identified in our study. Therefore, we do not include the LHAASO data in our fitting (See appendix \ref{sec:lhaaso} where we test the LHAASO model with HAWC data).
Based on the arguments outlined above, we use a pion decay model to explain the gamma-ray emission, similar to the approach in \cite{pionSNR}. We set the maximum energy values to 65 TeV and 1 PeV for the proton distribution for the test. The proton spectrum is modeled with both a broken power law and a power law with an exponential cutoff:

\begin{equation}\label{eq:sbpl}
f(E) = 
\begin{cases}
A \left( \dfrac{E}{E_0} \right)^{-\alpha_1}, & \text{for } E < E_{\rm break} \\
A \left( \dfrac{E_{\rm break}}{E_0} \right)^{\alpha_2 - \alpha_1} \left( \dfrac{E}{E_0} \right)^{-\alpha_2}, & \text{for } E \geq E_{\rm break}
\end{cases}
\end{equation}

\begin{equation}\label{eq:ecpl}
	f(E) = A(E/E_0)^{-\alpha} \exp(-E/E_{\rm cutoff}),
\end{equation}

where $A$ is the normalization of the spectrum at energy $E_0$; $\alpha_1$ and $\alpha_2$ are the spectral indices before and after the break energy $E_{\rm break}$, while $\alpha$ is the spectral index in the exponential cutoff model; $E_{\rm cutoff}$ is the cutoff energy. Since the target material needs to be considered, we assume an effective gas density $n$ of 20~cm$^{-3}$. This is the same value used in \cite{pionSNR}, which is estimated by considering the mass of the shocked gas ($\sim10^3\,M_{\odot}$) and the size of the remnant which ($\sim10$~pc). Furthermore, \cite{spectroIC443} made measurements which are consistent with this value. %similar as it was done in \cite{MWGSNR}. }
The fitted parameters results after running the aforementioned models are shown in Table \ref{tab:hadronicModel} together with the constraints that we use. The values of these constrains are similar to the ones in \cite{MWGSNR}, except for the spectral indices since the HAWC data suggest that they can be larger than 3.

\begin{deluxetable*}{cccccc}[!htbp]
\tablewidth{0pt}
\tablecaption{Parameter results of the Naima modeling. Two spectral functions for the cosmic-ray spectrum and two maximum proton energies were tested.\label{tab:hadronicModel}}
\tablehead{
\colhead{} & Model Constraints &\multicolumn{2}{c}{Broken Power Law} & \multicolumn{2}{c}{Exponential Cutoff} \\
\colhead{Parameter} & &\colhead{65 TeV} & \colhead{1 PeV} & \colhead{65 TeV} & \colhead{1 PeV}
}
\startdata
$\log(A\,[\rm TeV^{-1}])$ & 40-60 & $48.4 \pm 0.1$ & $48.4 \pm 0.1$ & $50.6 \pm 0.03$ & $50.6 \pm 0.03$ \\
$\alpha_1$ & 1.5-5 & $2.27 \pm 0.04$ & $2.28 \pm 0.04$ & $2.34 \pm 0.03$ & $2.35 \pm 0.03$ \\
$\alpha_2$ & 1.5-5 &  $3.01 \pm 0.06$ & $3.03 \pm 0.05$ & --- & --- \\
$E_{\rm break/cutoff}\,[\rm TeV]$ & 0.001-1000 & $0.10 \pm 0.04$ & $0.10 \pm 0.03$ & $1.2\pm0.2$ & $1.2^{+0.3}_{-0.2}$ \\
%$n\,[\rm cm^{-3}]$ & 0-1000 & $204^{+45}_{-36}$ & $252^{+68}_{-57}$ & $332^{+110}_{-91}$ & $290^{+102}_{-70}$ \\
BIC & --- &  65.1 & 65.2 & 111.2 & 111.2 \\
\enddata
\tablecomments{Uncertainties are statistical only.}
\end{deluxetable*}

The results are consistent when we compare between the different maximum energies in each of the cosmic-ray distributions. When comparing between spectral models, the broken power-law model provides a better fit than the exponential cutoff based on the BIC.  

\begin{figure*}
	\centering
	\plottwo{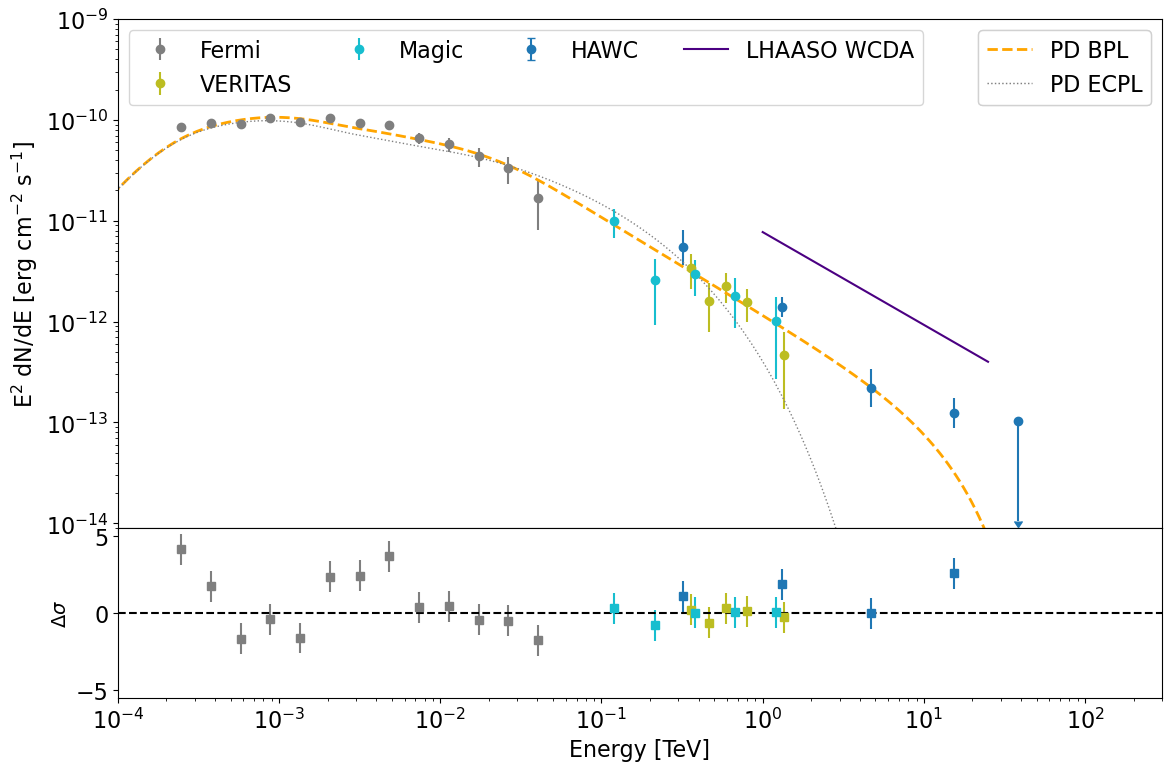}{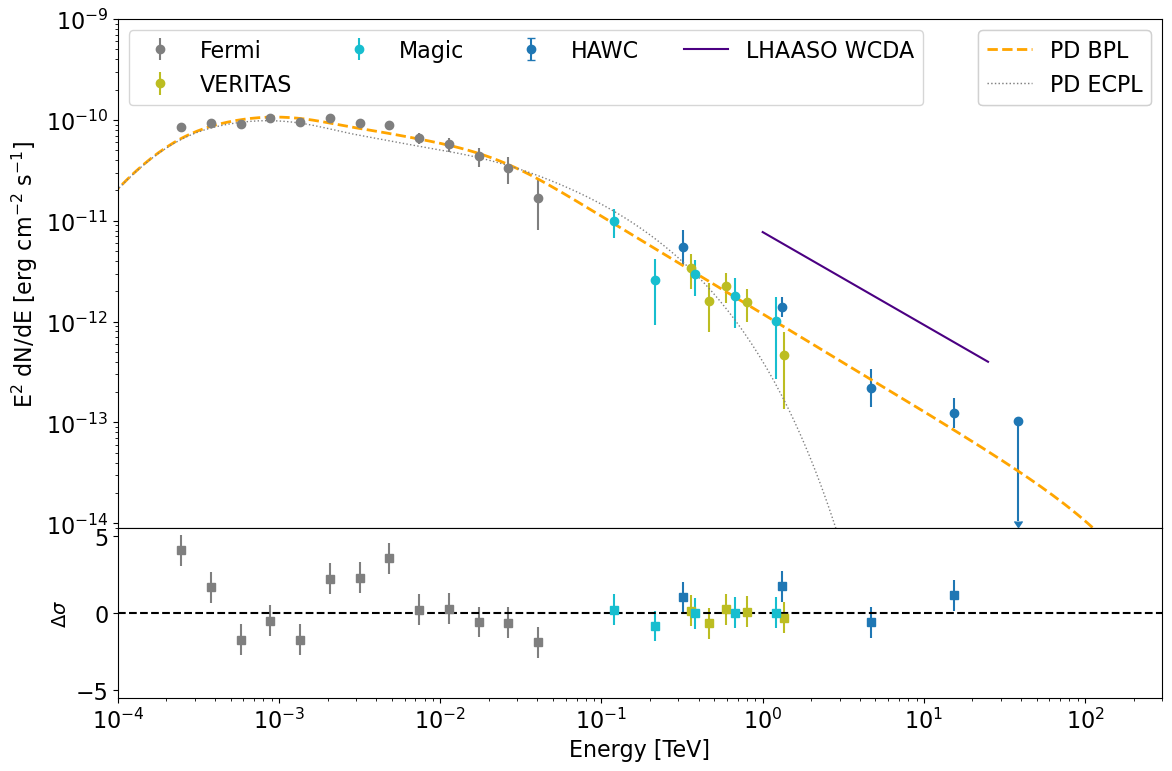}
	\caption{Spectrum of IC 443 in gamma rays. The model is a pion decay with a parent proton spectrum following a broken power law (BPL) or exponential cutoff spectrum (ECPL). On the left, the maximum energy of the protons is 65~TeV, on the right it is 1 PeV. These values are based on theory approximations as mentioned in the main text. In both cases the ECPL does not describe the data above 10 TeV. The model with a higher maximum proton energy provides a better fit to the tail of the spectral distribution.}\label{fig:hadronicIC443}
\end{figure*}

These models yield a total proton energy content with proton energies $E\geq0.8$~GeV of $W_p\sim [6-8]\times10^{49} (n/20 {\rm \,cm}^{-3})^{-1}$ erg, which is less than 1 per cent of the total supernova remnant energy of $\sim10^{51}$~erg, consistent with \cite{pionSNR}. Figure \ref{fig:hadronicIC443} shows the spectral distributions and the best Naima models. As can be see, a model with a higher maximum proton energy, in the case of the broke power law model, can describe better the tail of the gamma-ray spectrum.

\begin{figure}
	\centering
	\plotone{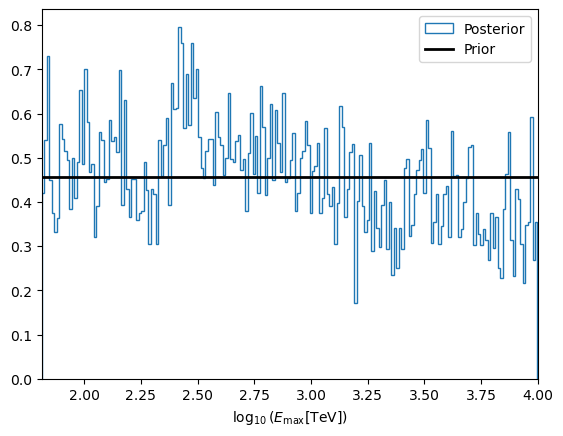}
	\caption{Prior and Posterior distribution of the maximum energy of the proton distribution after fitting a hadronic model with Naima. The distribution is still relatively flat compared to the prior distribution which is a uniform distribution in log space between 65 TeV and 10 PeV.}\label{fig:postEmax}
\end{figure}

We tested a model where we left the maximum energy of the proton distribution to be free. Since the HAWC data do not show any cutoff, the posterior distribution of the maximum energy remains relatively flat, as it is shown in Fig.~\ref{fig:postEmax}, compared to a prior uniform distribution in logarithmic space in the range of 65 TeV to 10 PeV.

This suggests that IC 443 has the potential to accelerate cosmic rays to energies beyond the ~65 TeV limit predicted by DSA, making it a strong candidate for a PeVatron. Given the constraints on leptonic processes, our findings further support IC 443 as a compelling source of PeV cosmic rays. However, higher-energy observations are still needed to determine where the gamma-ray spectrum cuts off, as this will provide a direct measure of the maximum energy achievable by protons.

It is important to remark that the spectral energy distribution shows a clear continuation of the spectrum observed by Fermi, Veritas and Magic, which can be taken as an indication that the gamma-ray emission observed by HAWC comes from the remnant. However, the maximum hotspot observed by HAWC is close to the pulsar CXOUJ J061705.3+222212 and emission from the pulsar wind nebula should not be discarded. Now, if there is emission from the pulsar wind nebula, this would come mostly from inverse Compton and it won't be able to explain the emission observed at the highest energies due to the Klein-Nishina effect. Currently, the angular resolution of HAWC cannot distinguish where precisely the emission is coming from.

\subsection{Origin of the gamma-ray emission from the extended source}

The extended emission we found in the analysis is surprising. First, it does not spatially correlate with the other supernova remnant G189.6, which is 1.2$^{\circ}$ away from the centroid of the extended source. We see no significant emission from this other remnant, which means that it does not emit gamma rays at a level detectable by HAWC. 
After discarding this possibility, we investigate three other scenarios next.

\subsubsection{Cosmic ray illumination of interstellar gas}
The first scenario is illumination of the surrounding interstellar gas by the accelerated cosmic rays from IC 443. Several studies, \citep[e.g.][]{modelMixedXRay}, show that IC 443 is embedded and interacting with the gas clouds surrounding the shell. We want to know if the cosmic rays from the IC 443 can escape and travel further away from the shell.

We start by estimating the power needed to explain the gamma-ray emission. Using the observed spectrum of the extended source and assuming that the gas is a few tens of parsecs away from IC 443 and a distance to the region of 1.5 kpc \citep[same as in][]{modelMixedXRay}, we calculate a gamma-ray luminosity of $\sim$3.17 $\times 10^{33}$ erg s$^{-1}$. If we assume that the centroid of the extended source is also located at the same distance as IC 443, we can use the width of the 2D Gaussian model to estimate a volume of this region. Assuming also a density of 1 cm$^{-3}$, and using Eq. 2 of \cite{energyDensCR}, we can estimate the energy density of the cosmic rays producing the gamma-ray emission for the extended source above $E_p = 10$ TeV ($E_{\gamma} = 5$ TeV), giving a value of 1.53 eV cm$^{-3}$. Using the volume again, we can estimate the energy budget of the cosmic rays which is  $\sim 1\times 10^{49}$ erg, lower than the total energy of the supernova remnant of $\sim10^{51}$ erg. However, we still need to determine whether there is target material in this region with which the cosmic rays can interact with. 

\begin{figure*}[!htbp]
	\centering
    \plottwo{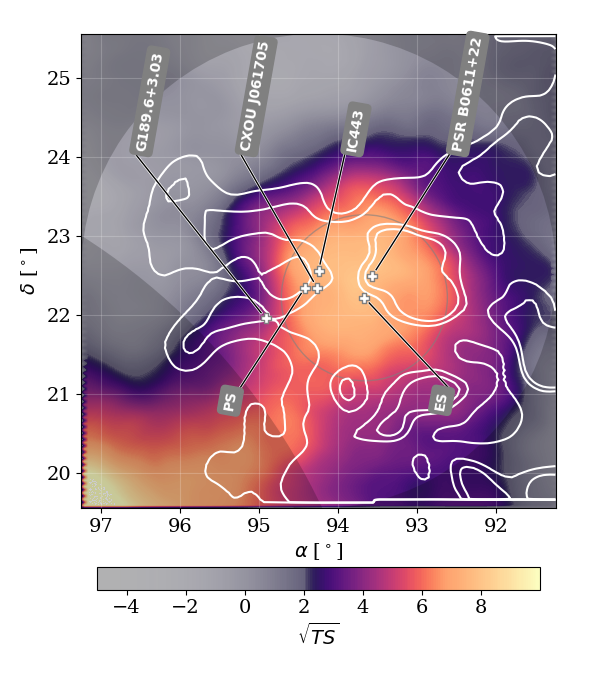}{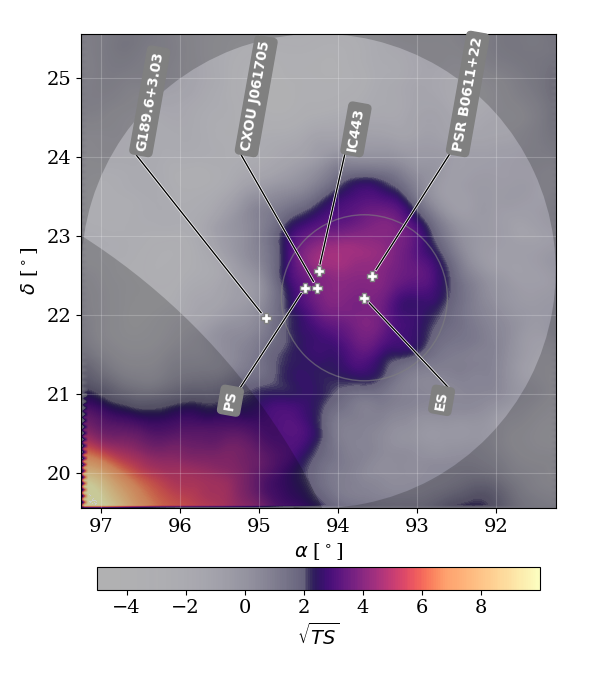}
	\caption{Left: The map is the same as the second map in Fig. \ref{fig:residualIndSources}, but now shows the contour lines of gas in the region from the HI4PI survey. The contour lines correspond to the column density determined after integrating the HI4PI data in the velocity range 3.5 to 10 km s$^{-1}$. Right: Residual map of the region after subtracting the model that includes the GDE, the point source and the gas template from the figure on the left. There is still unexplained emission left in the region.}\label{fig:residualIndSources2}
\end{figure*}

The gas surveys used in the GDE model are from the Dame~\citep{damesurvey} and the HI4PI~\citep{hi4pi} surveys. The GDE model considers the sea of cosmic rays interacting with the gas to produce pionic gamma rays and considers the whole gas in the line of sight. With this hypothesis, we test the possibility that IC 443 injects ``fresh" cosmic rays into the region as a second component besides the sea of cosmic rays, enhancing the production of pionic gamma rays. 
Using the kinematic distance tool of \cite{kinDist}, we estimate a range of velocities, going from 3.5 to 10 km s$^{-1}$ corresponding to distances of $\sim$0.8 to $\sim$2.97 kpc. We do not find significant molecular gas in the region that overlaps with the extended source. However, we find some gas in the atomic hydrogen survey that overlaps spatially with the extended emission. Figure~\ref{fig:residualIndSources2} shows the contour lines of the gas column density in the region from the HI4PI survey.

We repeated the analysis with a template model from the gas survey as the new morphology for the extended source \citep[similar to][]{gmcHAWC}. We assume a simple power-law function for the spectrum. We get a similar result as the 2D Gaussian model, with a normalization of $(3.88^{+1.5}_{-1.4})\times10^{-13}$~TeV$^{-1}$~cm$^{-2}$~s$^{-1}$ and index of $-2.48^{+0.1}_{-0.5}$. However, looking at the residual map in Fig.~\ref{fig:residualIndSources2}, we still see some emission in the region at the level of 4$\sigma$ that is not described by this model. A comparison of the significance histograms of the residual maps between the gas template and the model with the 2D gaussian source morphology is shown in Appendix \ref{sec:gasComp}.

\subsubsection{Faint or unresolved sources}
A second scenario is the possibility that the emission is the total contribution of faint (or unresolved) sources. Generally, when a gamma-ray telescope measures the flux of an extended source, it is possible that a component of that flux is from unresolved sources. This could be the case of the measurements of the GDE and the ES found in this analysis.
Based on the study in \cite{resUnresSources,resUnresSources2}, an estimate of the contribution of unresolved sources was calculated to be around $\sim$$28\%$ in the part of the Galaxy of our region of interest. The authors used data from LHAASO and HAWC to do their estimations. This would mean that there would still be some emission left that is not described by these faint sources.

However, the study used a simplified model of the Galaxy and it could be possible that a more dense profile of sources is present in this region. Nevertheless, we did not find in the literature any other source population in the region other than the pulsar B0611+22. A systematic study with this possibility will be part of a future publication focusing solely on this new extended source.

\subsubsection{A new TeV Halo}
The third hypothesis we investigate for the extended emission is a TeV halo. In Fig.~\ref{fig:residualIndSources}, the pulsar B0611+22 is in the middle of the extended source (0.29$^{\circ}$ from the best measured position). 

Middle-age pulsars ($\sim$100 kyr) seem to be the main sources of TeV (or electron) halos. These objects were first described by \cite{halos1} as a new type of source, mainly emitting gamma rays through inverse Compton radiation by the accelerated leptons in the PWN. Since then, new models and definitions have been developed for TeV halos. For example, \cite{halos2} mention that a halo forms after the pulsar and the pulsar wind nebula have escaped the remnant environment, which occurs at a time $\gtrsim100$ kyr, close to the age of B0611+22. According to this model, the energy density of the electrons in the region must be lower than the energy density of the interstellar medium, so that inverse Compton losses dominate over synchrotron and bremsstrahlung. 

We proceed to calculate some estimates that suggest that this emission is a potentially new TeV Halo.

Using the 2D Gaussian morphology results from Sect.~\ref{sec:results}, we can estimate the maximum diffusion coefficient in the region.
The upper limit for the diffusion coefficient is \citep{rdiffEq}
\begin{equation}\label{eq:DiffCoeff}
	D \leq \frac{R^2}{4t_e},
\end{equation}
where $t_e$ is the electron cooling time of 12 kyr at 100 TeV by upper-scattering photons from the CMB.  Using the gaussian width result, $\theta=1.05^{\circ}$, we can estimate a radius of this region based on geometry

\begin{equation}\label{eq:radius}
	R = d_0 \tan \theta,
\end{equation}
where $d_0 =3.55$ kpc, is the distance to the pulsar.

Using these calculations we get
\begin{equation}\label{eq:diffFhit}
	D \leq 2.21\times10^{28}\left(\frac{d}{d_0}\right)^2\left(\frac{t}{t_e}\right)^{-1} \text{cm}^2 \text{s}^{-1}.
\end{equation}
For this emission to be considered a halo, the maximum diffusion coefficient should be smaller than the diffusion coefficient of the ISM, which is $10^{30}$ cm$^2$ s$^{-1}$. We find that the diffusion coefficient to be $2.21\times10^{28}$ cm$^2$ s$^{-1}$, making this source a possible TeV Halo.

We also estimate the energy density of the electrons, $\epsilon_e$, and compare it to that of the interstellar medium. There are a couple of ways to calculate this. We use the pulsar properties to calculate $\epsilon_e = E_{inj}/V$, where $E_{\rm inj} = \dot{E}\tau_c$ is the injected energy, with $\dot{E}$ as the present spin-down power and $\tau_c$ as the characteristic age of the pulsar. To determine the volume, we use the radius estimated in Eq. \ref{eq:radius}. With $\dot{E}=6.24\times10^{34}$ erg s$^{-1}$ and $\tau_c=90$ kyr, this estimate gives $\epsilon_e\simeq0.003$ eV cm$^{-3}$.
  
\begin{figure*}
    \centering
	\plotone{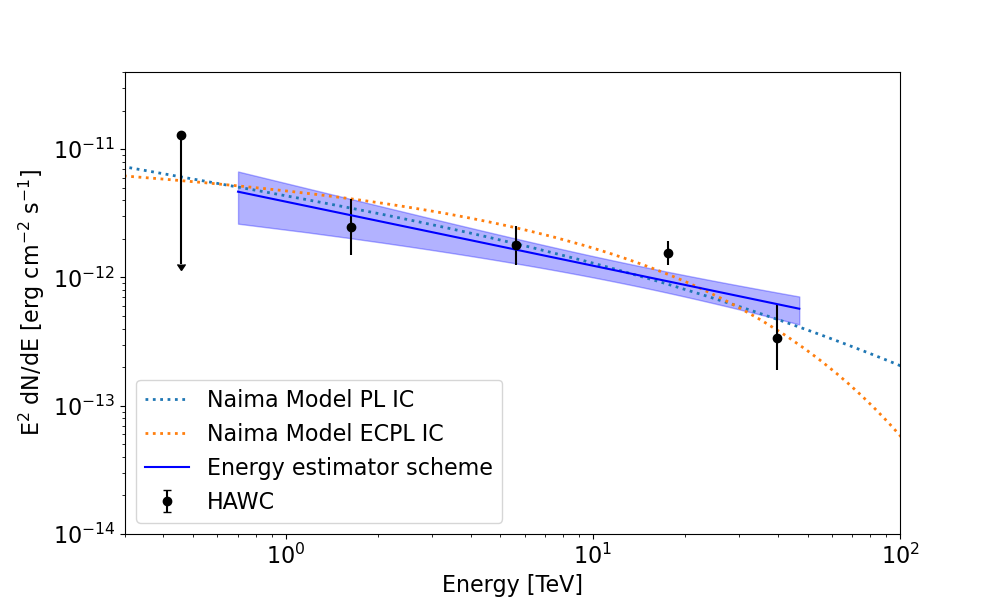}
	\caption{Spectrum of the extended source. A physical model assuming only inverse Compton is tested. Both simpler power law and exponential cutoff models for the electron spectrum can describe the data. The results of the parameters are used to estimate the electron energy density. For comparison, we show the power law fit to the HAWC data using the energy estimator scheme.}
	\label{fig:icModel}
\end{figure*}

We calculate a second estimate by evaluating the properties of the electron spectrum after fitting an inverse Compton model to the data as it is shown in Fig.~\ref{fig:icModel}. Integrating the simple power-law spectrum above 100 GeV yields an energy density of 0.0016 eV cm$^{-3}$, while for the exponential cutoff model, we obtain 0.0002 eV cm$^{-3}$. The parameters used in these models are presented in Appendix \ref{sec:ICHalo}. All estimates are below $\sim$0.1 cm$^{-3}$, the energy density of the interstellar medium. This provides some evidence that the newly observed emission may be associated with a TeV halo. Further investigation into a more comprehensive physical model to describe the emission will be the focus of future work.

\subsection{Supernova Remnant G189.6+03.3}\label{sec:g189p6}
G189.6+03.3 is another supernova remnant that is seen in the region of IC 443. As mentioned in the introduction, \cite{xrayG189IC443} mentioned that this remnant could be interacting with IC 443, which puts it at a similar distance as IC 443. The age of this supernova is older than IC 443, at least as reported by \cite{rosatIC443}, when it was reported in X-rays for the first time. 
In the Fermi Catalog of Extended Sources \citep{FermiCatExt}, the source FGES J0619.6+2229 is considered to be associated with G189.6+03.3. They found that it has a hard spectrum with an index of -2.0. This is intriguing since the HAWC data does not show significant emission in the region. To study this, we calculate the 95\% confidence level upper limits in the HAWC data. For this calculation we added a new source in our best fit model. This new source properties are obtained after we looked at Fermi data first.

We performed an analysis using \texttt{fermipy}. After fitting the sources in the 4FGL catalog, we found the excess that corresponds to G189.6+03.3 and modeled its morphology as a gaussian function and its flux as a power law.  We found the best position of the source to be $\alpha=(95.15\pm0.06)^{\circ}$, $\delta=(22.53\pm0.09)^{\circ}$ with a width of $(0.74\pm0.07)^{\circ}$. For the spectrum, we found an index of $-1.82\pm0.07$ and a normalization of $(9.39\pm1.90)\times10^{-13}$~MeV${-1}$~cm${-2}$~s${-1}$ at 1000~MeV and also calculated spectral points (See Appendix \ref{sec:appg189p6} for the configuration used).

\begin{figure*}
	\centering
	\plotone{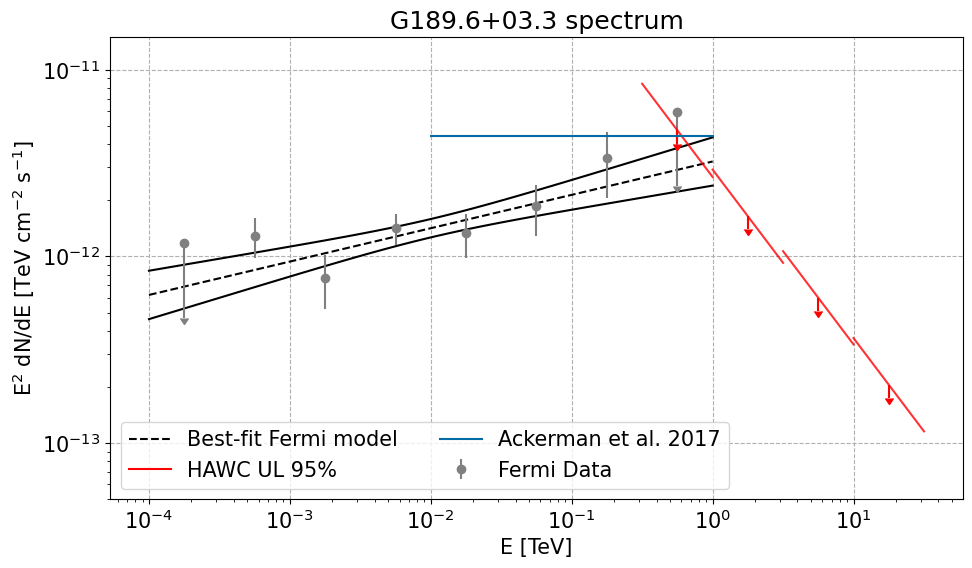}
	\caption{Spectrum of G189.6+03.3. Shown are the spectral points from Fermi data and upper limits from HAWC data. The limits suggest that the emission above 1 TeV is suppressed below HAWC's sensitivity.}
	\label{fig:ulsG189p6}
\end{figure*}

We then added the source in our best fit model of the HAWC analysis as a 2D gaussian morphology with a simple power law for the flux. We left all the parameters fixed except for the normalization. For the index we assume a value of -3.0, since based on the significance map, we expect the flux to fall.  
We then calculated the upper limits with HAWC by using the same method as in \cite{gmcHAWC}. The upper limits can be found in Table \ref{tab:ulsG189p6}. Fig. \ref{fig:ulsG189p6} shows that the flux from G189.6+03.3 above 1 TeV should drop off quickly, going below HAWC's sensitivity. 

\begin{deluxetable*}{cc}[!htbp]
\tablewidth{0pt}
\tablecaption{The 95\% C.L. upper limits using HAWC data of the remnant G189.6+03.3.\label{tab:ulsG189p6}}
\tablehead{
\colhead{Energy Range} & Upper limits  \\
\colhead{[TeV]} & [TeV$^{-1}$ cm$^{-2}$ s$^{-1}$]
}
\startdata
0.31 - 1 & 1.5e-11 \\
1 - 3.16 & 5.2e-13 \\
3.16 - 10 & 1.9e-14 \\
10 - 31.6 & 6.5e-16 \\
\enddata
%\tablecomments{}
\end{deluxetable*}

%
%_______________________________________________________________
\section{Conclusions}\label{sec:conclusion}
The new observations from HAWC in the region of IC 443 reinforce the importance of continuing to study this supernova remnant. The question of whether IC 443 is a PeVatron remains open. As discussed, if protons are indeed accelerated to PeV energies within this remnant, they could produce a significant gamma-ray contribution above 20 TeV, supporting the possibility of IC 443 as a PeVatron candidate. HAWC observations estimate the maximum proton energy to be above the theoretical minimum of 65 TeV, with no evidence of a cutoff in the gamma-ray spectrum, suggesting that further observations are essential to explore the potential for acceleration to PeV energies. 

No significant emission was detected from G189.6+03.3. The 95\% confidence level HAWC upper limits show that the flux of this source must fall quickly above 1 TeV.

While HAWC has also now detected an extended source in the region, which we called HAWC J0615+2213, with a centroid near the pulsar B0611+22, initial analyses indicate that cosmic-ray interactions with nearby gas cannot fully account for this emission. Estimates in Sect.~\ref{sec:discussion}, provide evidence that this extended source may represent a new TeV halo, although the scenario of unresolved sources needs to also be tested. Observations from LHAASO could help confirm this new source. Future experiments like SWGO could enhance the detection of additional halos in regions beyond HAWC’s current reach.

%% Please use the acknowledgment and contribution environments. This will 
%% be anonomyized when the "anonymous" style option is used. 
\begin{acknowledgments}
We acknowledge the support from: the US National Science Foundation (NSF); the US Department of Energy Office of High-Energy Physics; the Laboratory Directed Research and Development (LDRD) program of Los Alamos National Laboratory; Consejo Nacional de Ciencia y Tecnolog\'{i}a (CONACyT), M\'{e}xico, grants LNC-2023-117, 271051, 232656, 260378, 179588, 254964, 258865, 243290, 132197, A1-S-46288, A1-S-22784, CF-2023-I-645, c\'{a}tedras 873, 1563, 341, 323, Red HAWC, M\'{e}xico; DGAPA-UNAM grants IG101323, IN111716-3, IN111419, IA102019, IN106521, IN114924, IN110521 , IN102223; VIEP-BUAP; PIFI 2012, 2013, PROFOCIE 2014, 2015; the University of Wisconsin Alumni Research Foundation; the Institute of Geophysics, Planetary Physics, and Signatures at Los Alamos National Laboratory; Polish Science Centre grant, 2024/53/B/ST9/02671; Coordinaci\'{o}n de la Investigaci\'{o}n Cient\'{i}fica de la Universidad Michoacana; Royal Society - Newton Advanced Fellowship 180385; Gobierno de España and European Union-NextGenerationEU, grant CNS2023- 144099; The Program Management Unit for Human Resources \& Institutional Development, Research and Innovation, NXPO (grant number B16F630069); Coordinaci\'{o}n General Acad\'{e}mica e Innovaci\'{o}n (CGAI-UdeG), PRODEP-SEP UDG-CA-499; Institute of Cosmic Ray Research (ICRR), University of Tokyo. H.F. acknowledges support by NASA under award number 80GSFC21M0002. C.R. acknowledges support from National Research Foundation of Korea (RS-2023-00280210). We also acknowledge the significant contributions over many years of Stefan Westerhoff, Gaurang Yodh and Arnulfo Zepeda Dom\'inguez, all deceased members of the HAWC collaboration. Thanks to Scott Delay, Luciano D\'{i}az and Eduardo Murrieta for technical support.
\end{acknowledgments}

%\begin{contribution}
%%This section gives authors the space to recognize author contributions. The text inside this environment is NOT counted towards the total word quanta. At a minimum, manuscripts are expected to include this text:

%All authors contributed equally to the Terra Mater collaboration.

%% But authors are expected to provide more specific details, e.g. 
%%
%%SC was responsible for writing and submitting the manuscript.
%%WWM came up with the initial research concept and edited the manuscript.
%%OTS obtained the funding and edited the manuscript.
%%EBF provided the formal analysis and validation. He also edited the manuscript.
%%GEH Supervised the undergraduates, wrote the software and administers the project github and Zenodo repositories.
%%
%% Authors can use the Contributor Role Taxonomy (CRediT) at
%% for ideas on how write a good statement tailored to their needs.

%\end{contribution}

\facilities{HAWC}

\software{Threeml \citep{threeml}, HAL \citep{hal}}
%\software{astropy \citep{2013A&A...558A..33A,2018AJ....156..123A,2022ApJ...935..167A},  
%          Cloudy \citep{2013RMxAA..49..137F}, 
%          Source Extractor \citep{1996A&AS..117..393B}
%          }


\appendix

\section{Morphology model}

\subsection{2D Gaussian in the sphere}

The general shape for the 2D gaussian is:
\begin{equation}
	    f(\vec{x}) = \left(\frac{180^\circ}{\pi}\right)^2 \frac{1}{2\pi \sqrt{\det{\Sigma}}}  \, {\rm exp}\left( -\frac{1}{2} (\vec{x}-\vec{x}_0)^\intercal \cdot \Sigma^{-1}\cdot (\vec{x}-\vec{x}_0)\right) 
\end{equation}

where : 

$\vec{x}_0 = ({\rm RA}_0,{\rm Dec}_0)$, 

$\Lambda = \left( \begin{array}{cc} \sigma^2 & 0 \\ 0 & \sigma^2 (1-e^2) \end{array}\right) $, 

$U = \left( \begin{array}{cc} \cos \theta & -\sin \theta \\ \sin \theta & cos \theta \end{array}\right)$,

$\Sigma = U\Lambda U^\intercal$

For a symmetric gaussian, $e=0$ and $\theta = 0$. 

\section{Model Selection Statistics}

The tables below show the log likelihood values, AIC and BIC values of the models that we try. Table \ref{tab:ps_sel}, shows the results from Algorithm \ref{alg:ps}. Table \ref{tab:es_sel} shows the results of Algorithm \ref{alg:es}. Table \ref{tab:spec_sel}, shows the results from Algorithm \ref{alg:spectrum}.

\begin{deluxetable}{ccccc}[!htbp]
\tablewidth{0pt}
\tablecaption{Statistical results from the search in Algorithm \ref{alg:ps}.\label{tab:ps_sel}}
\tablehead{\colhead{Model Description} & \colhead{-log Likelihood} & \colhead{AIC} & \colhead{BIC} & $TS_{\rm model}$}
\startdata
Fitting GDE & 48275.6 & 96553.3 & 96563.8 & 123.8 \\
GDE + 1 PS & 48253.9 & 96517.7 & 96570.1 & 43.6 \\
GDE + 2 PS & 48241.2 & 96496.4 & 96569.8 & 25.3 \\
GDE + 3 PS & 48234.7 & 96503.5 & 96503.5 & 13.0 \\
\enddata
\tablecomments{$TS_{\rm model}$ is calculated between rows. First $TS_{\rm model}$ is with respect the estimated background from HAWC. With 3 point sources, the $TS_{\rm model}$ is below the threshold established in Algorithm \ref{alg:ps}, so we stop the search. PS here means point source. }
\end{deluxetable}
The 2 PS correspond to the point source located at the HAWC counterpart of IC 443 and the centroid of the extended source in the main results. 

\begin{deluxetable}{ccccc}[!htbp]
\tablewidth{0pt}
\tablecaption{Statistical results from the search in Algorithm \ref{alg:es}.\label{tab:es_sel}}
\tablehead{\colhead{Model Description} & \colhead{-log Likelihood} & \colhead{AIC} & \colhead{BIC} & $TS_{\rm model}$}
\startdata
GDE + 1 ES + 1 PS & 48236.4 & 96484.9 & 96547.8 & 9.5 \\
GDE + 1 PS + 1 ES & 48232.7 & 96477.4 & 96540.3 & 17.0 \\
GDE + 1 PS + 1 ES All-parameter fit & 48229.9 & 96480.0 & 96584.9 & 22.6 \\
\enddata
\tablecomments{$TS_{\rm model}$ is calculated by comparing the log likelihood of the 2-point source model to the new model with an extended source. The position of the sources are fixed, except for the last model after we found the best morphology of the region.  }
\end{deluxetable}
The first model, corresponds to the one where we changed the highest TS source, in this case IC 443, to an extended source. As seen by the table, the second model where IC 443 is a point source together with an extended source, is the model that crosses the threshold that we set up in Algorithm \ref{alg:es}.

\begin{deluxetable}{ccccc}[!htbp]
\tablewidth{0pt}
\tablecaption{Statistical results from the search in Algorithm \ref{alg:spectrum}.\label{tab:spec_sel}}
\tablehead{\colhead{Model Description} & \colhead{-log Likelihood} & \colhead{AIC} & \colhead{BIC} & $TS_{\rm model}$}
\startdata
GDE + 1 PS (PL) + 1 ES (PL) & 48229.8 & 96469.6 & 96522.0 & -- \\
GDE + 1 PS (LP) + 1 ES (PL) & 48229.9 & 96471.8 & 96534.8 & -0.3 \\
GDE + 1 PS (PL) + 1 ES (LP) & 48224.4 & 96460.8 & 96523.7 & 10.8 \\
GDE + 1 PS (CO) + 1 ES (PL) & 48230.0 & 96472.0 & 96534.9 & -0.4 \\
GDE + 1 PS (PL) + 1 ES (CO) & 48224.2 & 96460.4 & 96523.3 & 11.2 \\
\enddata
\tablecomments{In all the models, we only fit the spectral parameters. First row is the baseline to calculate the $TS_{\rm model}$. As can be seen, there's not enough evidence to suggest that there's a curvature in the spectrum in the sources. We tried a log parabola (LP) and a exponential cutoff (CO) spectra. }
\end{deluxetable}

\color{black}
\section{Comparison with LHAASO measurements}\label{sec:lhaaso}

\begin{figure*}[!htbp]
	\centering
	\plottwo{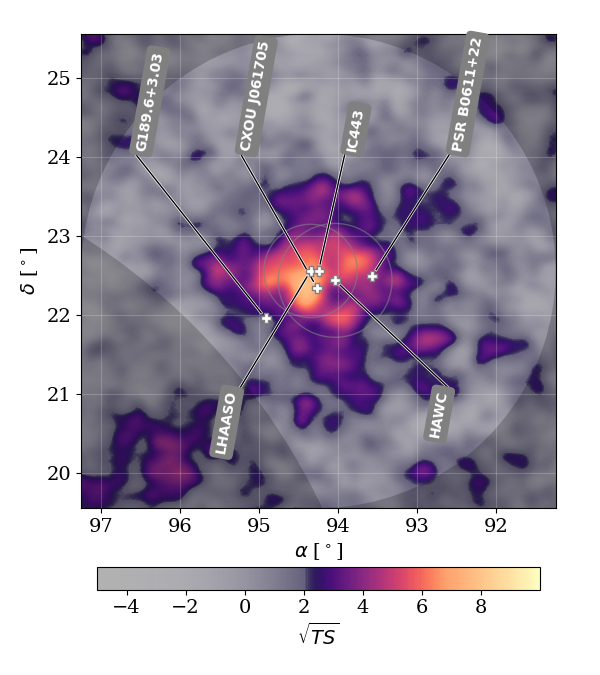}{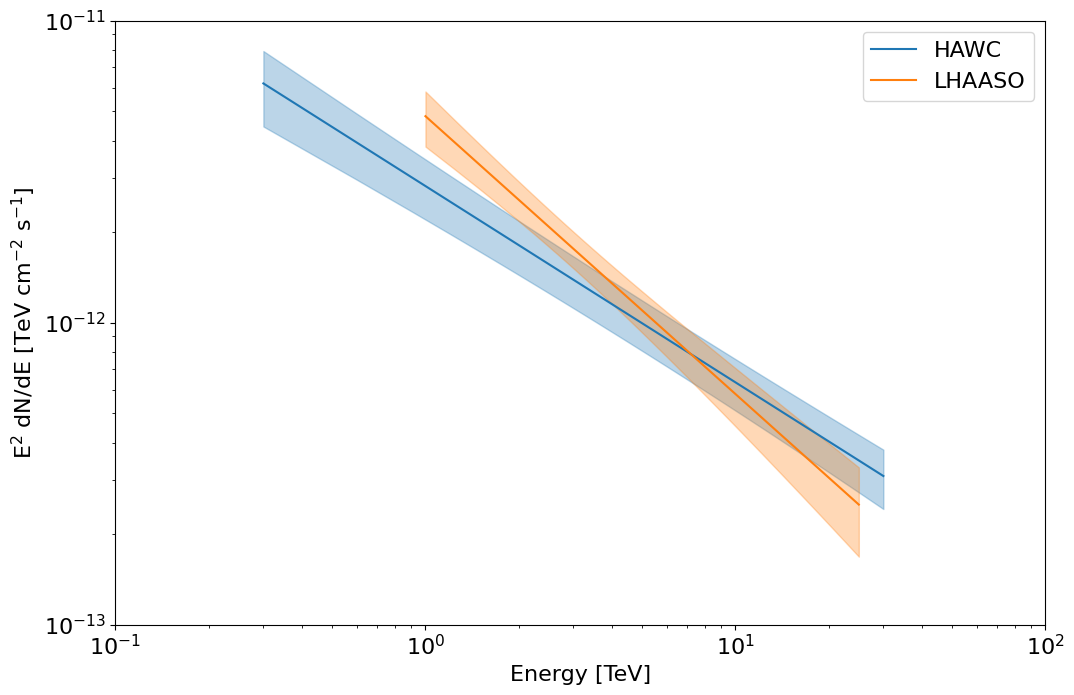}
	\caption{Left: Sky map with the positions measured by LHAASO and HAWC using only a 2D Gaussian morphology. Circles are the 1$\sigma$ of the Gaussian function. Right: SED of the extended source. Fluxes are consistent within uncertainties.  }\label{fig:comparison}
\end{figure*}

The LHAASO collaboration presented their catalog results in \cite{lhaasoCat}. For the catalog search, in the region of IC 443 they found an extended sources, modeled with a 2D Gaussian function for the morphology, and a simple power-law function for the spectrum. The energy range where they significantly detect the source is between 1 TeV to 25 TeV.
Using the same model, we refit the region of IC 443 using the $f_{\rm hit}$ scheme to compare results. One thing to consider is that in the LHAASO catalog, the galactic diffuse emission model is based on the Planck maps of dust optical depth \citep{planck}. 
Table \ref{tab:comparison},  and Fig. \ref{fig:comparison} shows the results, which are consistent within uncertainties.
\textcolor{black}{Table \ref{tab:comparison2} shows the statistic comparison between models. Based on the table we can tell that there is some evidence that the model with a point source and an extended source is slightly better to the one extended source model.}
\begin{deluxetable}{ccc}[!htbp]
\tablewidth{0pt}
\tablecaption{\textcolor{black}{Comparison of the fitted parameters} between HAWC and LHAASO in the region of IC 443 using the same model \textcolor{black}{with one extended source}.\label{tab:comparison}}
\tablehead{
\colhead{Parameter} & \colhead{LHAASO} & \colhead{HAWC}
}
\startdata
R.A. [deg] & $94.35 \pm 0.18$ & $94.04 \pm 0.12$ \\
Decl. [deg] & $22.57 \pm 0.18$ & $22.44 \pm 0.12$ \\
$\sigma$ [deg] & $0.59 \pm 0.08$ & $0.72 \pm 0.10$ \\
$\Phi_0$ & $1.95 \pm 0.27$ & $1.54^{+0.34}_{-0.28}$ \\
$\alpha$ & $-2.92 \pm 0.14$ & $-2.65 \pm 0.07$ \\
GDE* & --- & $4.0 \pm 0.9$ \\
\enddata
\tablecomments{Uncertainties are statistical only. $\Phi_0$ is in units of $10^{-13}\,\mathrm{TeV^{-1}\,cm^{-2}\,s^{-1}}$ and evaluated at a pivot energy of 3~TeV. *The GDE scale factor is unitless.}
\end{deluxetable}

\begin{deluxetable}{ccc}[!htbp]
\tablewidth{0pt}
\tablecaption{Statistics comparison between the one-extended-source model (LHAASO's model) and the best fit model found in our analysis with HAWC data in the region of IC 443.\label{tab:comparison2}}
\tablehead{
\colhead{Parameter} & \colhead{Best Fit Model} & \colhead{One extended source model}
}
\startdata
$-\ln{L}$& 48229.9 & 48239.5 \\
$TS_{\rm model}$ & 91.4 & 72.2 \\
AIC & 96479.9 & 96491.0 \\
BIC & 96584.8 & 96554.0 \\
\enddata
\tablecomments{$TS_{\rm model}$ is calculated with respect to the GDE-only model which has a $-\ln{L}=48275.6$.}
\end{deluxetable}

\section{Comparison of significance histograms}
\label{sec:gasComp}
Significance histograms of the residual maps using the models where the extended source is described by a 2D Gaussian, and when it is described by the gas template. 
The 2D Gaussian model has a distribution better described by the expectation. 
\begin{figure*}[!htbp]
	\centering
    \plottwo{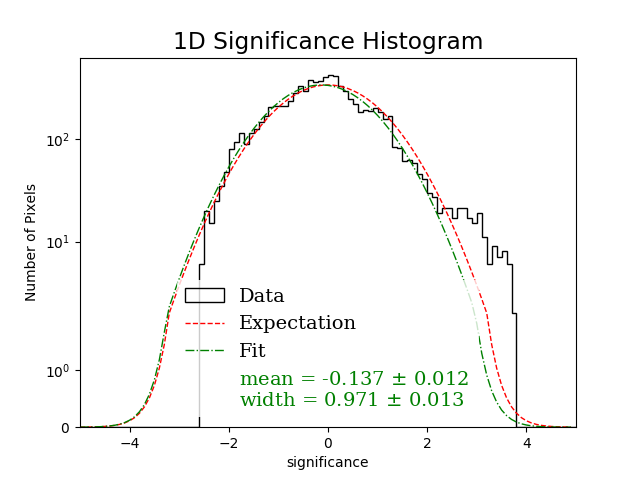}{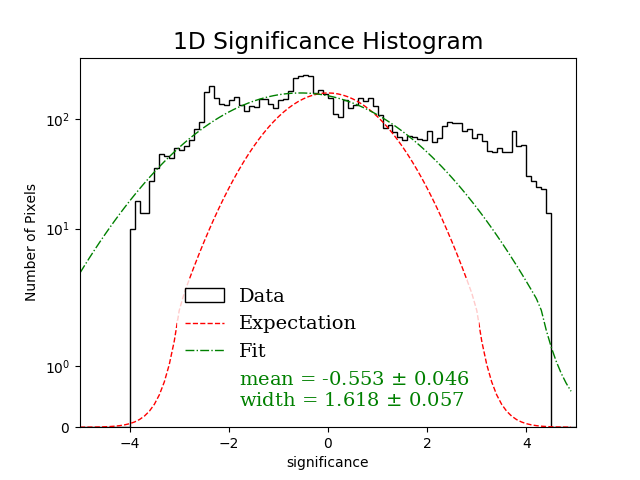}
	\caption{Significance histograms of the residual maps. First histogram is from the map where the extended sources is described by a 2D Gaussian. Second histogram is where the extended source is described by the gas template. First histogram is closer to what we expect from only background emission.} 
	\label{fig:sigComp}
\end{figure*}
 
\section{Parameter results of the inverse Compton modeling of the extended source}\label{sec:ICHalo}

We used Naima for the modeling. The electron spectra used when fitting the gamma-ray flux of the extended source are a simple power law and a power law with exponential cut off. For the inverse Compton model, we used the default photon fields that exist in Naima, which include CMB, near-infrared and far-infrared. The temperatures of 2.72 K, 30 K, and 3000 K, and energy densities of 0.261, 0.5, and 1 eV cm$^{-3}$ are used respectively for each field \citep{naima}. 
The results of these models are:

\begin{itemize}
	\item Simple power law: Normalization: $(5.3^{+3.9}_{-2.5})\times 10^{47}$ erg$^{-1}$; index: 3.4$\pm$0.2.
	\item Power law with exponential cutoff: Normalization: $(1.3^{+1.7}_{-1.0})\times 10^{47}$ erg$^{-1}$; index: 2.9$^{+0.4}_{-0.7}$; energy cutoff: $67^{+241}_{-44}$ TeV.
\end{itemize}

\section{Fermi Analysis of G189.6+03.3}\label{sec:appg189p6}
We performed a binned likelihood analysis of Fermi-LAT data using the Fermipy package. The region of interest was centered at R.A. = 94.25$^{\circ}$, Dec. = 22.57$^{\circ}$, with a width of 6$^{\circ}$ and a spatial bin size of 0.05$^{\circ}$ in celestial coordinates. The analysis used 8 energy bins per decade spanning the energy range from 100 MeV to 1 TeV. We selected Pass 8 SOURCE class events (evclass = 128) and included both FRONT and BACK converting events (evtype = 56), applying a zenith angle cut of 90$^{\circ}$ to suppress contamination from Earth limb photons. The time range of the data extended from MET 239557417 to 730300274. The instrument response function used was \texttt{P8R3\_SOURCE\_V3}, and energy dispersion corrections were enabled for all components except the isotropic and Galactic diffuse templates. The diffuse background model included the Galactic interstellar emission model gll\_iem\_v07.fits and the isotropic template iso\_P8R3\_SOURCE\_V3\_v1.txt. We included all 4FGL catalog sources within 10$^{\circ}$ of the ROI center. Source extension tests were performed with radial sizes ranging from 0.00316$^{\circ}$ to 1.0$^{\circ}$, sampled logarithmically in 26 steps.

%% For this sample we use BibTeX plus aasjournalv7.bst to generate the

\bibliography{biblio.bib}
\bibliographystyle{aasjournalv7}



\end{document}